\documentclass[trackchanges, twocolumn]{aastex701} 

\usepackage{amsmath}

\usepackage[normalem]{ulem}
\usepackage{xcolor}

\newif\ifshowcomments
\showcommentstrue          

\begin{document}

\title{J-PAS: Semi-Supervised Sim-to-Obs Transfer for Robust Star--Galaxy--Quasar Classification}

\author[orcid=0000-0002-0279-4874,gname=Daniel,sname={L\'{o}pez-Cano}]{Daniel L\'{o}pez-Cano}
\affiliation{Department of Mathematical Physics, Institute of Physics, University of São Paulo, R. do Matão 1371, 05508-090, São Paulo, SP, Brazil}
\affiliation{Fundação de Amparo à Pesquisa do Estado de São Paulo (FAPESP), R. Pio XI, 1500, Alto da Lapa, 05468-901, São Paulo, SP, Brasil}
\email[show]{daniellopezcano13@gmail.com}

\author[orcid=0000-0001-8295-7022, gname=L.\ Raul, sname=Abramo]{L.\ Raul Abramo}
\affiliation{Department of Mathematical Physics, Institute of Physics, University of São Paulo, R. do Matão 1371, 05508-090, São Paulo, SP, Brazil}
\email{raulabramo@usp.br}

\author[orcid=0000-0001-6480-1155, sname=L. Nakazono]{L. Nakazono}
\affiliation{Observatório Nacional, Rua General José Cristino, 77, São Cristóvão, 20921-400 Rio de Janeiro, RJ, Brazil}
\email{lilianne.nakazono@on.br}

\author[orcid=0000-0001-6979-0125]{I. Pérez-Ràfols}
\affiliation{Departament de F\'{i}sica, EEBE, Universitat Polit\`{e}cnica de Catalunya, c/Eduard Maristany 10, 08930 Barcelona, Spain}
\email{ignasi.perez.rafols@upc.edu}

\author[orcid=0000-0002-8812-1135]{G. Martínez-Solaeche}
\affiliation{Instituto de Astrofísica de Andalucía (IAA-CSIC), P.O. Box 3004, 18080 Granada, Spain}
\email{gimarso@iaa.es}

\author[orcid=0000-0002-9553-4261]{J. Chaves-Montero}
\affiliation{Institut de Física d'Altes Energies (IFAE), The Barcelona Institute
of Science and Technology, 08193 Bellaterra (Barcelona), Spain}
\email{jchaves@ifae.es}

\author{Matthew M. Pieri}
\affiliation{Aix Marseille Univ, CNRS, CNES, LAM, Marseille, France}
\email{matthew.pieri@lam.fr}

\author{Jailson Alcaniz}
\affiliation{Observatório Nacional, Rua General José Cristino, 77, São Cristóvão, 20921-400 Rio de Janeiro, RJ, Brazil}
\email{}

\author{Narciso Benitez}
\affiliation{Independent Researcher}
\email{}

\author{Silvia Bonoli}
\affiliation{Donostia International Physics Center (DIPC), Manuel Lardizabal Ibilbidea, 4, San Sebastián, Spain}
\email{}

\author{Saulo Carneiro}
\affiliation{Observatório Nacional, Rua General José Cristino, 77, São Cristóvão, 20921-400 Rio de Janeiro, RJ, Brazil}
\email{}

\author{Javier Cenarro}
\affiliation{Centro de Estudios de Física del Cosmos de Aragón (CEFCA), Plaza San Juan, 1, E-44001, Teruel, Spain}
\affiliation{Unidad Asociada CEFCA-IAA, CEFCA, Unidad Asociada al CSIC por el IAA y el IFCA, Plaza San Juan 1, 44001 Teruel, Spain}
\email{}

\author{David Cristóbal-Hornillos}
\affiliation{Centro de Estudios de Física del Cosmos de Aragón (CEFCA), Plaza San Juan, 1, E-44001, Teruel, Spain}
\email{}

\author{Simone Daflon}
\affiliation{Observatório Nacional, Rua General José Cristino, 77, São Cristóvão, 20921-400 Rio de Janeiro, RJ, Brazil}
\email{}

\author{Renato Dupke}
\affiliation{Observatório Nacional, Rua General José Cristino, 77, São Cristóvão, 20921-400 Rio de Janeiro, RJ, Brazil}
\email{}

\author{Alessandro Ederoclite}
\affiliation{Centro de Estudios de Física del Cosmos de Aragón (CEFCA), Plaza San Juan, 1, E-44001, Teruel, Spain}
\email{}

\author{Rosa González Delgado}
\affiliation{Instituto de Astrofísica de Andalucía - CSIC, Apdo 3004, E-18080, Granada, Spain}
\email{}

\author{Antonio Hernán-Caballero}
\affiliation{Centro de Estudios de Física del Cosmos de Aragón (CEFCA), Plaza San Juan, 1, E-44001, Teruel, Spain}
\affiliation{Unidad Asociada CEFCA-IAA, CEFCA, Unidad Asociada al CSIC por el IAA y el IFCA, Plaza San Juan 1, 44001 Teruel, Spain}
\email{}

\author{Carlos Hernández--Monteagudo}
\affiliation{Instituto de Astrofísica de Canarias, C/ Vía Láctea, s/n, E-38205, La Laguna, Tenerife, Spain}
\affiliation{Universidad de La Laguna, Avda Francisco Sánchez, E-38206, San Cristóbal de La Laguna, Tenerife, Spain}
\email{}

\author{Jifeng Liu}
\affiliation{National Astronomical Observatory of China, Chinese Academy of Sciences, Beijing, China}
\email{}

\author{Carlos López-Sanjuan}
\affiliation{Centro de Estudios de Física del Cosmos de Aragón (CEFCA), Plaza San Juan, 1, E-44001, Teruel, Spain}
\affiliation{Unidad Asociada CEFCA-IAA, CEFCA, Unidad Asociada al CSIC por el IAA y el IFCA, Plaza San Juan 1, 44001 Teruel, Spain}
\email{}

\author{Antonio Marín-Franch}
\affiliation{Centro de Estudios de Física del Cosmos de Aragón (CEFCA), Plaza San Juan, 1, E-44001, Teruel, Spain}
\affiliation{Unidad Asociada CEFCA-IAA, CEFCA, Unidad Asociada al CSIC por el IAA y el IFCA, Plaza San Juan 1, 44001 Teruel, Spain}
\email{}

\author{Claudia Mendes de Oliveira}
\affiliation{Departamento de Astronomia, Instituto de Astronomia, Geofísica e Ciências Atmosféricas, Universidade de São Paulo, Brazil}
\email{}

\author{Mariano Moles}
\affiliation{Centro de Estudios de Física del Cosmos de Aragón (CEFCA), Plaza San Juan, 1, E-44001, Teruel, Spain}
\email{}

\author{Fernando Roig}
\affiliation{Observatório Nacional, Rua General José Cristino, 77, São Cristóvão, 20921-400 Rio de Janeiro, RJ, Brazil}
\email{}

\author{Laerte Sodré Jr.}
\affiliation{Departamento de Astronomia, Instituto de Astronomia, Geofísica e Ciências Atmosféricas, Universidade de São Paulo, Brazil}
\email{}

\author{Keith Taylor}
\affiliation{Instruments4, 4121 Pembury Place, La Canada Flintridge, CA 91011, U.S.A.}
\email{}

\author{Jesús Varela}
\affiliation{Centro de Estudios de Física del Cosmos de Aragón (CEFCA), Plaza San Juan, 1, E-44001, Teruel, Spain}
\email{}

\author{Héctor Vázquez Ramió}
\affiliation{Centro de Estudios de Física del Cosmos de Aragón (CEFCA), Plaza San Juan, 1, E-44001, Teruel, Spain}
\affiliation{Unidad Asociada CEFCA-IAA, CEFCA, Unidad Asociada al CSIC por el IAA y el IFCA, Plaza San Juan 1, 44001 Teruel, Spain}
\email{}

\author{Jose Vilchez}
\affiliation{Instituto de Astrofísica de Andalucía - CSIC, Apdo 3004, E-18080, Granada, Spain}
\email{}

\author{Javier Zaragoza-Cardiel}
\affiliation{Centro de Estudios de Física del Cosmos de Aragón (CEFCA), Plaza San Juan, 1, E-44001, Teruel, Spain}
\affiliation{Unidad Asociada CEFCA-IAA, CEFCA, Unidad Asociada al CSIC por el IAA y el IFCA, Plaza San Juan 1, 44001 Teruel, Spain}
\email{}


\begin{abstract}
Modern studies in astrophysics and cosmology increasingly rely on simulations and cross-survey analyses, yet differences in data generation, instrumentation, calibration, and unmodeled physics introduce distribution mismatches between datasets (``domain shift''). In machine-learning pipelines, this occurs when the joint distribution of inputs and labels differs between the training (source) and application (target) domains, causing source-trained models to underperform on the target. Transfer learning and domain adaptation provide principled ways to mitigate this effect.
We study a concrete simulation-to-observation case: semi-supervised domain adaptation (SSDA) to transfer a four-class spectral classifier---high-redshift quasars, low-redshift quasars, galaxies, and stars---from J-PAS mock catalogs based on DESI spectra to real J-PAS observations. Our pipeline pretrains on abundant labeled DESI$\rightarrow$J-PAS mocks and adapts to the target domain using a small labeled J-PAS subset. We benchmark SSDA against two baselines: a J-PAS--only supervised model trained with the same target-label budget, and a mocks-only model evaluated on held-out J-PAS data.
On this held-out J-PAS data, SSDA achieves a macro-F1 score (balancing precision and recall) of $0.82$ and an overall true positive rate of $0.89$, compared to $0.79/0.85$ for the J-PAS--only baseline and $0.73/0.87$ for the mocks-only model. The gains are driven primarily by improved quasar classification, especially in the high-redshift subclass ($\mathrm{F1}=0.66$ vs.\ $0.55/0.37$), yielding better-calibrated candidate lists for spectroscopic targeting (e.g., WEAVE-QSO) and AGN searches.
This study shows how modest target supervision enables robust, data-efficient simulation-to-observation transfer when simulations are plentiful but target labels are scarce.
\end{abstract}




\section{Introduction}\label{sec:intro}

\setcounter{footnote}{0}


Astronomical analyses increasingly aim to combine datasets that share a common goal but differ in project-specific characteristics. On the survey side, many current/forthcoming missions such as the Dark Energy Spectroscopic Instrument~\citep[DESI,][]{2025JCAP...02..021A}, Euclid~\citep{2025A&A...697A...1E}, the Legacy Survey of Space and Time~\citep[LSST,][]{2019ApJ...873..111I}, WEAVE~\citep{2016SPIE.9908E..1GD, 2024MNRAS.530.2688J}, or J-PAS~\citep{2014arXiv1403.5237B}, will observe overlapping sky regions with different passbands, depths, footprints, and calibration strategies. 
Analysing them together to perform cross-survey analyses can effectively enlarge survey area and redshift reach, improve completeness, and enable stronger systematic testing. At the same time, heterogeneity in selection, observing conditions, and processing pipelines makes cross-survey work nontrivial.

Simulations constitute another pillar in modern cosmological analyses. For example, $N$-body runs model the formation and evolution of haloes and of the large-scale structure of the Universe \citep[e.g.,][]{2016MNRAS.463.2273F, 2017ComAC...4....2P, 2021MNRAS.508..575G, 2021MNRAS.506.2871S, 2025JCAP...02..020R}, hydrodynamical simulations describe galaxy formation and feedback processes \citep[e.g.,][]{2014MNRAS.444.1518V, 2015MNRAS.450.1937C, 2023MNRAS.526.4978S}, and strong lensing pipelines can generate mock images resembling direct observations \citep[e.g.,][]{2023A&A...674A..79B, 2023MNRAS.518.2746A}. These tools can be employed to conduct survey validation checks, produce end-to-end forecasts,  and train machine-learning (ML) pipelines to perform different tasks. However, rendering any of these simulation outputs into realistic instrument-level observables requires making several assumptions (selection, noise, blending, reduction, etc.), and small differences at any level can accumulate, so simulated mocks and real data typically exhibit some degree of mismatch in properties and characteristics.

Taken together, survey heterogeneity and the approximations required to translate simulations into realistic observations mean that any model trained in one setting will face a different data--label relationship in another~\citep[e.g.][]{2025arXiv250703086P}. We refer to this mismatch as \emph{domain shift}: the joint distribution over inputs and labels differs between the \emph{source} (training) and \emph{target} (application) domains. 
As a result, the performance of a given model, when evaluated over the target domain, will typically show reduced accuracy, worse calibration, and biased downstream results.


A practical set of techniques has recently been developed by the ML community to mitigate these effects~\citep[see][ for a recent review]{2022arXiv220807422L}.
\emph{Domain adaptation (DA)} methods transfer knowledge from a labelled source to a related target domain under the same task but different input distributions.


{The practical regimes are largely set by target-label availability: with many target labels available, it's common to employ supervised transfer techniques \citep[e.g.][]{2024arXiv240314608H}; with none, \emph{unsupervised domain adaptation (UDA)} aligns source/target representations \citep[e.g.,][]{2015arXiv150507818G, 2017arXiv170205464T, 2017arXiv170508584L}; and with a small labelled target subset, \emph{semi-supervised DA (SSDA)} uses those few labels to improve model calibration \citep[e.g.,][]{2021arXiv210604732B, 2023arXiv230202335Y}.}

{Related approaches include \emph{domain generalization} -- for robustness across multiple sources without seeing the target \citep[see][for recent reviews]{2021arXiv210302503Z, 2021arXiv210303097W} and \emph{contrastive (CL)/self-supervised learning (SSL)} -- for determining domain-agnostic embeddings from unlabeled data, both of which can improve invariance to nuisance variations \citep[e.g.,][]{2020arXiv200205709C, 2021arXiv210504906B, 2025arXiv250416929A}.}

{These approaches have already proved useful in astronomy and cosmology (see \citealt{2023RASTI...2..441H} for a review of CL/SSL): feature-alignment improves robustness to changing depth/PSF in image-based strong lensing tasks \citep[e.g.,][]{2023ApJ...954...28A, 2023arXiv231117238S, 2024arXiv241016347S, 2024arXiv241103334A, 2025ApJ...990...47P, 2025MLS&T...6c5032P}; DA/CL reduce bias and improve coverage in simulation-based inference, and enhance the performance of forward model frameworks \citep[e.g.,][]{2023arXiv231101588R, 2024MNRAS.527.7459A, 2024ApJ...975...38L, 2024A&A...685A..37L, 2025Ap&SS.370...14A, 2025arXiv250805744A}; and DA/DG/SSL improve label efficiency and robustness in cross-survey classification and photometric-redshift pipelines \citep[e.g.,][]{2021ApJ...911L..33H, 2021MNRAS.506..677C, 2023MLS&T...4b5013C, 2024ApJ...961...51V, 2024Astro...3..189G}.}

{Overall, these results support a simple message: domain shift is widespread in astrophysical pipelines, and domain-aware learning can improve robustness, calibration, and reproducibility when transferring models across surveys or from simulations to observations.}

In this work, we apply one of these domain-shift-aware techniques---in particular, a Semi-Supervised Domain Adaptation (SSDA) approach---to improve the performance of an ML model for classifying astronomical sources based on their J-PAS narrow-band flux vectors (commonly referred to as \emph{J-spectra}). The Javalambre Physics of the Accelerating Universe Astrophysical Survey \citep[J-PAS,][]{2014arXiv1403.5237B} is an ongoing wide-field multiband survey providing quasi-spectroscopic photometry over several thousand square degrees (see \url{https://www.j-pas.org}). J-PAS employs $54$ contiguous narrow-band filters supplemented by two medium-band and one broad-band filter to deliver precise photometric redshifts for galaxies and quasars (QSOs), map stellar populations, and probe large-scale structure. In this work we use the J-PAS Early Data Release (EDR; \url{https://www.j-pas.org/datareleases/jpas_early_data_release}); J-PAS software and data products are maintained within the CEFCA infrastructure, including \url{https://gitlab.cefca.es}.

We focus on a concrete, relevant case: training a four-way classifier for \emph{J-spectra} data distinguishing high-redshift quasars, low-redshift quasars, galaxies, and stars. This problem is not merely methodological; accurate quasar probabilities directly enable operational candidate selection for spectroscopic follow-up, where false positives translate into wasted fibres and reduced science yield. In practice, target-selection programs require ranking sources at very low false-positive rates while preserving completeness for rare classes; this places a premium on well-calibrated probabilities rather than just hard labels. In this sense, improved quasar classification feeds directly into survey strategy: it increases the purity of high-$z$ candidate lists, stabilizes forecasts, and helps make the resulting selection function explicit for downstream LSS analyses.

We use the public J-PAS Early Data Release (EDR) as a testing ground for domain-shift--aware classification, leveraging public DESI EDR spectroscopy \citep{2024AJ....168...58D} together with DESI$\rightarrow$\emph{J-spectra} mocks built by projecting DESI spectra into the J-PAS narrow-band system \citep[following][]{2023MNRAS.520.3476Q}.
Source classification in J-PAS and J-PLUS has been widely explored using ML techniques \citep[e.g.,][]{2021A&A...645A..87B, 2024MNRAS.527.3347V, 2024A&A...691A.221D, 2025arXiv251120524J}, including the miniJPAS quasar-selection efforts for creating combined catalogs \citep{2025arXiv250711380P, 2023A&A...678A.144P, 2023A&A...673A.103M, 2023MNRAS.520.3494R, 2023MNRAS.520.3476Q}.
A recurring outcome of these pipelines is the presence of a \emph{simulation-to-observation} gap, whereby models trained primarily on mocks degrade when applied to real data.
Beyond methodology, accurate, calibrated class probabilities are operationally relevant for targeting in upcoming spectroscopic efforts such as \textsc{WEAVE} \citep{2016sf2a.conf..259P, 2024MNRAS.530.2688J}, where low false-positive rates at fixed completeness for rare high-$z$ quasars directly translate into improved fibre usage and clearer selection functions for downstream large-scale structure analyses.

We adopt a SSDA scheme: pretrain on DESI$\rightarrow$J-PAS mocks and adapt on J-PAS using a small labelled subset from the DESI$\times$J-PAS cross-match, so that mocks supply coverage while target labels guide class-conditional alignment.
We compare three settings on J-PAS: (i) \emph{SSDA} (this work), (ii) a \emph{target-only} supervised baseline using the same labelled J-PAS budget, and (iii) a \emph{mocks-only} model evaluated zero-shot on J-PAS (i.e., trained solely on DESI$\rightarrow$J-PAS mocks and applied directly to J-PAS observations with no fine-tuning, reweighting, or calibration using J-PAS sample).
Performance is quantified with class-resolved diagnostics, emphasising the quasar subclasses most consequential for follow-up targeting under domain shift.
We release the full pipeline at \url{https://github.com/daniellopezcano/JPAS_Domain_Adaptation}---data curation, training and SSDA adaptation code, sweep configurations, and plotting scripts--- including YAML files for the \texttt{wandb} sweeps and scripts to regenerate all tables and figures.

The remainder of the paper is organised as follows: Sect.~\ref{sec:data} describes the J-PAS data, the DESI$\rightarrow$J-PAS mocks, the DESI$\times$J-PAS cross-match, preprocessing, and the train/validation/test splits. Sect.~\ref{sec:methods} details the model, training regimes, and losses. Sect.~\ref{sec:results} presents the main results with class-resolved diagnostics (confusion matrices, F1, etc.). Sect.~\ref{sec:conclusions} summarizes our findings and discusses the implications for cross-survey and sim-to-obs pipelines.


\begin{figure}
  \centering
  \includegraphics[width=1.0\columnwidth]{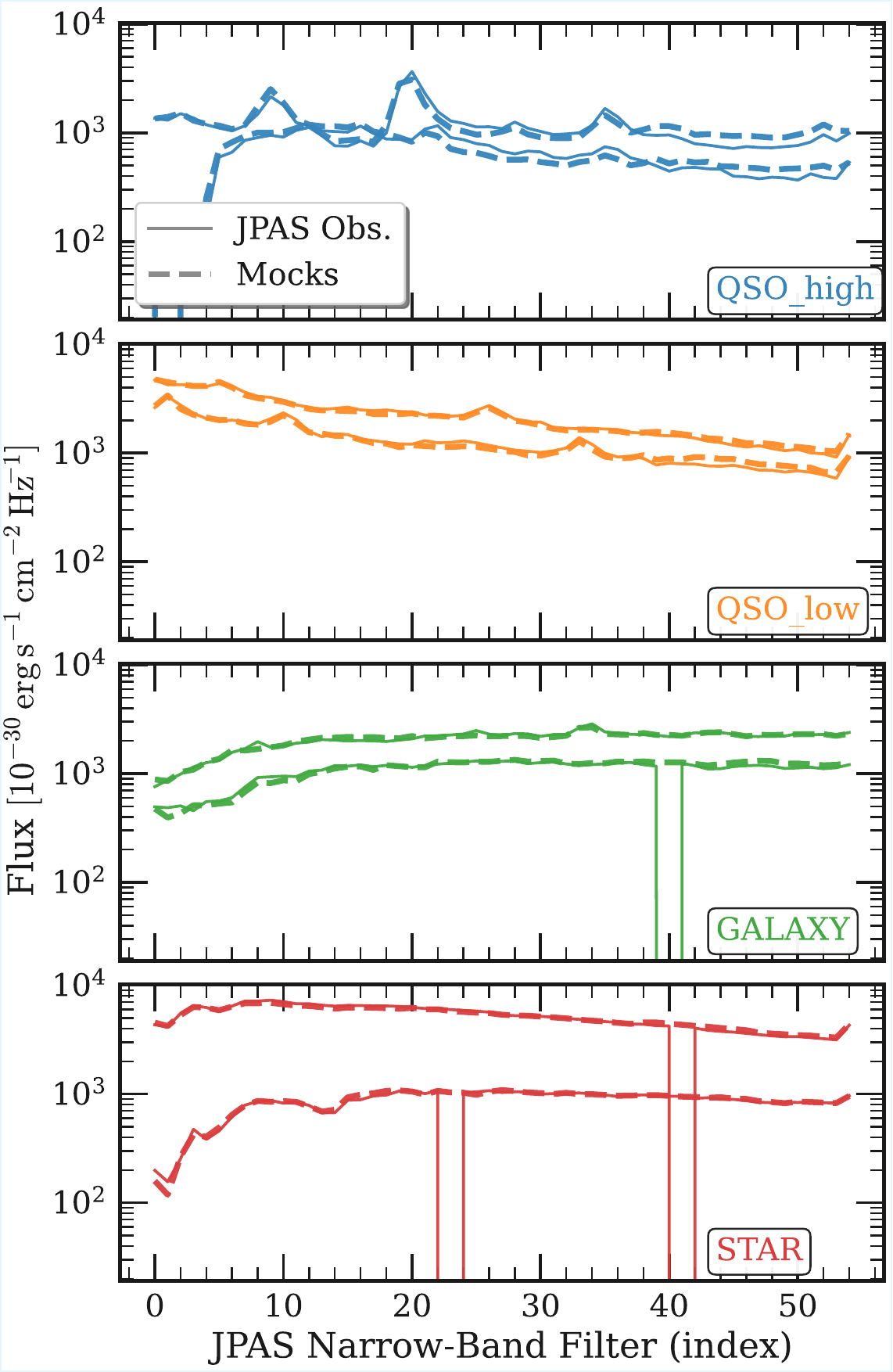}
  \caption{
    Representative photometric SEDs, comparing J-PAS (solid) with DESI$\rightarrow$J-PAS \emph{Mocks} (dashed).
    Each solid/dashed pair corresponds to a distinct individual object with a one-to-one mock counterpart (i.e., the dashed curve is the mock constructed for that same source); the multiple curves shown in each panel are therefore not averaged templates but separate objects plotted together for illustration.
    Panels correspond to different classes.
    Curves span the effective 55-band set after reliability
    masking (Section~\ref{subsec:datasets_curation}).
    For visual clarity, we display examples with $i$-band signal-to-noise ratio $>200$; this high-S/N cut is used only for visualization and is not applied elsewhere in the analysis.
    Within each class, the plotted objects are selected from the subset passing this visualization-only S/N cut (randomly chosen from the eligible set).
    For legibility, we also omit error bars. We intentionally retain bands with missing flux measurements in J-PAS (visible as gaps) to reflect real observational coverage.}
  \label{fig:data_spectra}
\end{figure}

\section{Data}\label{sec:data}

Our work focuses on classifying J-PAS sources into four distinct classes denoted as: high-redshift quasars (\texttt{QSO\_high}), low-redshift quasars (\texttt{QSO\_low}), galaxies (\texttt{GALAXY}), and stars (\texttt{STAR}).

We employ two different datasets to train and evaluate the performance of our models. 
First, we build a large labelled \emph{DESI$\rightarrow$J-PAS mocks} sample by projecting DESI spectra into the J-PAS photometric system. 
Full details on the procedure to generate simulated J-PAS fluxes from spectra were presented in \cite{2023MNRAS.520.3476Q}, with the key difference being that for the present paper, instead of using the SDSS spectra we now employ DESI spectra.

\subsection{DESI$\rightarrow$J-PAS mock generation}\label{subsec:mocks_generation}

Each DESI spectrum is convolved with the J-PAS transmission curves and passed through an observational model (zeropoints and band-dependent noise) to produce $54$ narrow-band J-PAS-like flux estimates. We intentionally exclude the two medium bands at the spectral edges of J-PAS, as the DESI spectroscopic wavelength coverage does not reliably span these passbands; generating mock fluxes there would require extrapolation and could introduce extra domain-dependent systematics. All experiments, therefore, use only the 54 narrow bands plus the $i$ band.

We then apply an \emph{object-by-object multiplicative normalization} to place the synthetic \emph{J-spectra} on the same absolute scale as imaging photometry: we compute synthetic broad-band fluxes from the DESI spectrum in an anchor band and rescale by the ratio of observed-to-synthetic flux. In our pipeline the anchors are the J-PAS $i$ band and the Legacy Surveys $r$ band (used to tie the DESI spectrophotometry to photometry). This step is purely multiplicative (no wavelength-dependent warping) and therefore does not, by construction, introduce color-dependent distortions. The per-band noise model for the mocks is anchored to the J-PAS flux uncertainties measured in the same photometric system (see below), so the resulting synthetic \emph{J-spectra} reproduce the signal-to-noise levels of the chosen J-PAS photometry.

This provides a synthetic data set of labeled observations with approximately $1.5\times10^6$ unique sources spanning the four classes with good coverage of rare classes such as \texttt{QSO\_high}.

Second, we assemble a \emph{J-PAS$\times$DESI cross-match} -- hereafter \emph{J-PAS Observations} -- that attaches DESI spectroscopically generated labels to J-PAS photometric sources {(from the Early Data Release catalogue)} in the \emph{target} measurement space. This set contains $52{,}020$ objects before quality cuts, with a strongly imbalanced composition (e.g., $\lesssim2\%$ in \texttt{QSO\_high}, similar to the mocks). While too small to train high-capacity models by itself, this dataset is ideal for calibration and for assessing performance where it ultimately matters: on real J-PAS observations.

For each J-PAS source, we extract per-band fluxes using the $3\arcsec$ aperture photometry, including the catalogue PSF aperture correction computed with the help of non-saturated high signal-to-noise ratio (SNR) point-like sources. This choice is the most appropriate for the predominantly unresolved sources, but it has proven the most robust overall in terms of providing the best estimate of the intrinsic SED in a consistent aperture across bands; the corresponding flux uncertainties from this same photometry are propagated into the mock-generation step to set the per-band noise and reproduce realistic J-PAS-like S/N. We apply catalogue-level quality filtering to remove problematic photometry before constructing the final working sets; the remaining curation steps are listed in Sect.~\ref{subsec:datasets_curation}.

Figure~\ref{fig:data_spectra} shows representative real \emph{J-spectra} (solid) with their corresponding DESI$\rightarrow$J-PAS \emph{mock} counterparts (dashed) for each class. Each data vector spans the effective 55-band set after reliability masking (see subsection~\ref{subsec:datasets_curation}) and is displayed using the same per-source normalization to facilitate visual comparison. As expected, stars show relatively smooth continua; galaxies exhibit color changes and break-like features; and the quasar subclasses (\texttt{QSO\_low}, \texttt{QSO\_high}) present localized modulations associated with strong emission features sampled by the narrow bands. The mocks track the overall shapes and relative fluxes well, while small, localized residuals in color and amplitude appear at specific filters and near the edges of the wavelength coverage. Occasional gaps in the J-PAS curves reflect bands removed by coverage/quality cuts, which the mocks do not inherit. This object-by-object view clarifies what the simulator reproduces faithfully (global SED structure and class-separating cues) and where residual mismatches concentrate (localized bands and missing-filter patterns). Even after careful emulation, the simulations show slight differences from the actual J-PAS observations, introducing a \emph{domain shift}.

Unless noted otherwise, each object in our datasets is further equipped with a one-hot \textit{The Tractor} morphology indicator \citep{2016ascl.soft04008L}. \textit{The Tractor} is a model-based photometry framework present in the DESI Legacy Surveys, which fits a compact set of parametric light-profile families to imaging data and records the best-fit class (e.g., PSF, GPSF, REX, EXP, DEV, SER, GGAL). These classifications are available for all objects in both J-PAS observations and DESI$\rightarrow$J-PAS mocks. The resulting discrete label provides a coarse size/shape cue---separating point-like from extended sources---and is particularly useful for disambiguating stars and QSOs from galaxies at low S/N in J-PAS. We encode this categorical field as a one-hot vector and concatenate it to the photometric features so that the classifiers can condition on morphology without overriding the spectral information.

In the remainder of this section we describe how the raw datasets are curated and organized for the classification experiments.

\subsection{Dataset curation and splits}\label{subsec:datasets_curation}

\begin{figure}
  \centering
  \includegraphics[width=1.0\columnwidth]{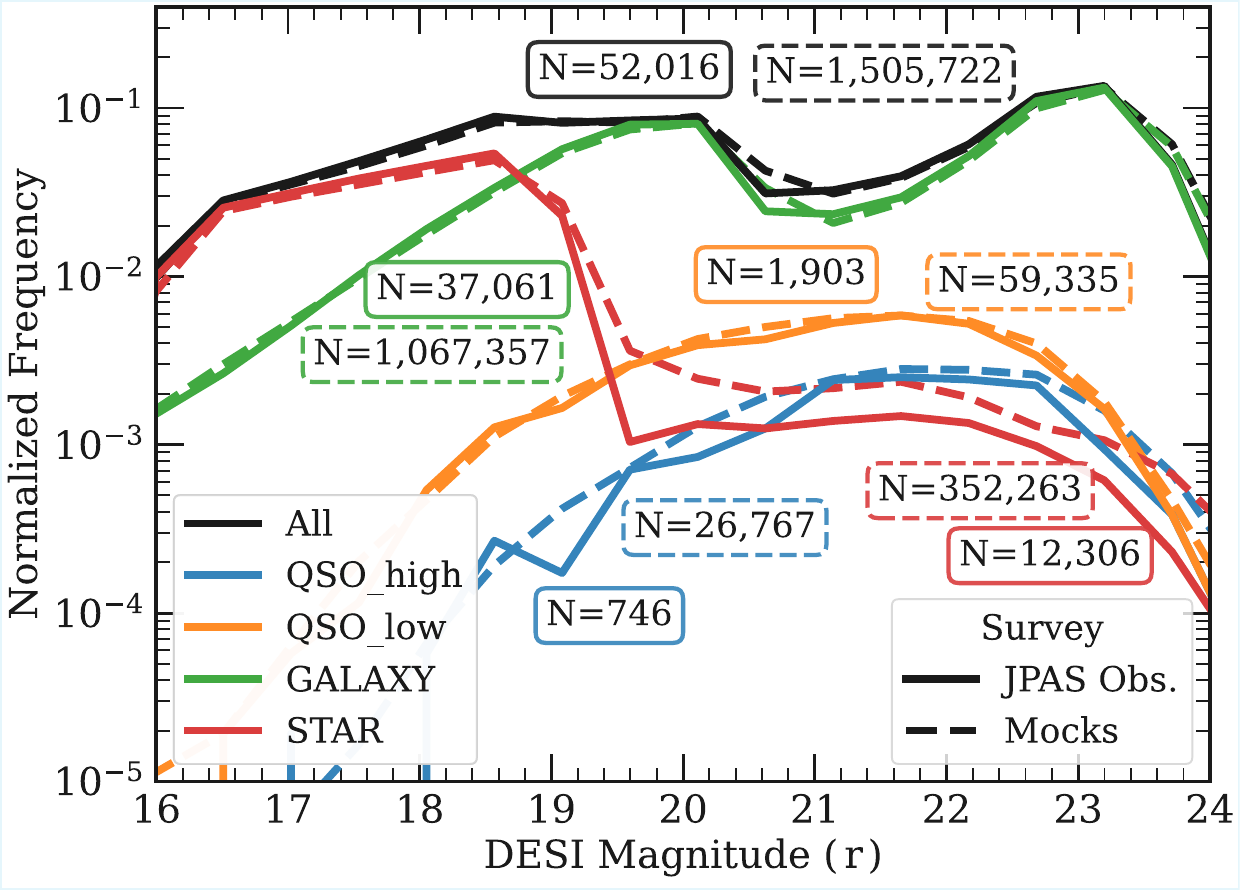}
  \caption{
    Class balance in our two working datasets:
    full DESI$\rightarrow$J-PAS \emph{Mocks} (source; dashed) versus the \emph{labeled}
    J-PAS$\times$DESI subset (target; solid).
    The y-axis shows \emph{relative frequency per magnitude bin} normalized by the
    total size of the mock catalog ($\simeq 1.5\times10^{6}$ unique sources); consequently,
    curves appear on the same percentage scale and trace each other in shape
    even though absolute counts differ widely (\,$N_{\mathrm{Mocks}}\!\gg\!N_{\mathrm{J\text{-}PAS}}$; see the total per-class counts in the annotated boxes).
    The figure includes \emph{all objects after our data curation} (Sect.~\ref{subsec:datasets_curation});
    no train/validation/test splits or magnitude cuts are applied here.
  }
  \label{fig:histogram_Mocks_Obs_class_distributions}
\end{figure}

As introduced in Sect.~\ref{sec:data}, our working datasets -- \emph{DESI$\rightarrow$J-PAS mocks} (source) and \emph{J-PAS obs.} (target) -- occupy the same photometric space after reliability masking: 54 J-PAS-like narrow-band filters plus the i-band measurements. Upon these two raw datasets, we apply the following curation steps:

\begin{enumerate}
    \item \textit{Missing and sentinel values.}
    We drop SED rows that are entirely missing (all entries NaN) or contain sentinel placeholders ($-99$ or $99$). No such rows are present in the \emph{J-PAS Obs.} set as constructed here, whereas the synthetic data occasionally include these placeholders from the mock-generation process; we remove $1{,}041$ fully-NaN SEDs and $609$ rows containing sentinel values in the mocks. For partially missing SEDs, we retain the object and linearly interpolate along the wavelength-ordered grid using the closest valid neighbors (affecting $\sim 1.18\%$ of mock SEDs). This interpolation is performed \emph{independently for each object} (per-SED): only the valid flux values of the same SED enter the interpolation. This preserves coverage without introducing target-driven imputation.

    \item \textit{Classes and class imbalance.}
    We classify sources into four classes: \texttt{QSO\_high} ($z\geq2.1$), \texttt{QSO\_low} ($z<2.1$), \texttt{GALAXY}, and \texttt{STAR}. 
    We adopt $z=2.1$ as an operational (survey-driven) threshold commonly used in large optical spectroscopic surveys to distinguish quasars whose spectra provide a usable Ly$\alpha$ forest from those that do not. For $z\gtrsim2.1$, the Ly$\alpha$ forest region blueward of $1216\,\AA$ is shifted into the ground-based optical window with sufficient path length for cosmological analyses. This value is therefore widely used in target selection and pipeline definitions of “high-$z$” (Ly$\alpha$-forest) quasars. We emphasize that this boundary does not mark a sharp physical transition in quasar properties, but rather reflects observational and analysis requirements.
    Figure~\ref{fig:histogram_Mocks_Obs_class_distributions} shows per-class counts as a function of magnitude in both domains. The relative abundances are closely matched overall but strongly imbalanced, especially for \texttt{QSO\_high}.

    Three takeaways follow from Fig.~\ref{fig:histogram_Mocks_Obs_class_distributions}: (i) the dependence of counts on magnitude reflects DESI selection inherited by both domains; (ii) relative abundances track closely between domains globally and per class; and (iii) the problem is strongly imbalanced, with \texttt{QSO\_high} comprising only $\approx$\,1.4--1.8\% overall and becoming rarer at bright magnitudes. This motivates our use of a class-aware objective (balanced cross-entropy; see Sect.~\ref{sec:methods}) and macro-averaged/class-resolved evaluation metrics (Sect.~\ref{sec:results}).
    
    \item \textit{Cross-match and de-duplication across domains (leakage prevention).}
    We construct a linking table between \emph{mocks} and \emph{J-PAS Obs.} using the \texttt{TARGETID} variable contained in both datasets\footnote{We construct the DESI$\times$J-PAS association with a nearest-neighbor sky match within a $1\arcsec$ radius, using great-circle separations on the sphere. When multiple DESI targets fall within $1\arcsec$, we keep the closest match and discard ambiguous cases.
    }.
    Conceptually, this enables `leakage-safe' splits (see point four below):
    Any mock object linked to a J-PAS source assigned to a target validation/test split is excluded from the source training pool; conversely, J-PAS objects used for training/validation (e.g., in the JPAS-supervised baseline or the small labelled set for SSDA) are excluded from the J-PAS test set. Thus, the same astrophysical source never influences training and evaluation in a different disguise (synthetic vs.\ observed), and target metrics reflect genuine sim$\rightarrow$obs generalization rather than memorization via duplicates.
    
    \item \textit{Train/validation/test splits (and evaluation mask).}
    For \emph{DESI$\rightarrow$J-PAS mocks} (source) we adopt a 70/15/15 split with stable class fractions, removing any overlaps with the J-PAS cross-match before splitting to prevent leakage. For \emph{J-PAS Obs.} (target) we split 30/30/40 into train/validation/test to preserve statistical power during evaluation for sparse regimes, especially \texttt{QSO\_high}. The target \emph{J-PAS Obs.} training set is therefore limited to $\sim 1.5\times10^{4}$ objects, which motivates the use of SSDA rather than target-only training. After defining these splits, we apply an evaluation-only magnitude mask: in the test sets of both domains, we retain only sources with $r \le 22.5$ (our scientific target operating range), while training and validation remain untrimmed to leverage additional faint examples and stabilize rare-class learning.
    
    \item \textit{Normalization.}
    We compute per-band standardization statistics only on the source training split (mocks) and then apply the resulting fixed transform to both domains. For band $b$ with training mock mean $\mu_b$ and standard deviation $\sigma_b$, each flux $x_b$ is transformed to $(x_b-\mu_b)/\sigma_b$. A single, source-derived normalization enforces a shared scale while preventing the target distribution from influencing feature scaling.

\end{enumerate}

In summary, after performing basic dataset curation and leak-safe splits, both the source \emph{mocks} and the labelled \emph{J-PAS observations}, live in the same input space: 55 JPAS-like bands plus a one-hot Tractor morphology flag. The training and validation splits will be employed for the different experiments described in  Section~\ref{sec:methods}, and the test samples will be used for a fair evaluation comparison in Section~\ref{sec:results}.

\section{Methods}\label{sec:methods}

The data setup in Sect.~\ref{sec:data} leads naturally to our training plan: we have a large, label-rich \emph{source} (DESI$\rightarrow$J-PAS mocks) and a smaller but measurement-faithful \emph{target} (J-PAS Observations), both curated to share the same feature space. In what follows we define the loss used to handle class imbalance, and compare three complementary training regimes for four-way classification: (i) a mocks-only model evaluated zero-shot on J-PAS, (ii) a semi-supervised domain adaptation (SSDA) variant that adapts source-pretrained features with a small labeled J-PAS subset, and (iii) a J-PAS-supervised baseline trained only on target labels. We also summarize the hyperparameter sweeps that establish model convergence.

\subsection{Loss function: balanced cross-entropy}\label{subsec:balanced_ce}

All models are trained with a multi-class \emph{balanced cross-entropy} (BaCE) loss. Let $\mathbf{Y}\!\in\!\{0,1\}^{N\times C}$ be the true-label targets for a batch of size $N$ over $C{=}4$ classes and $\hat{\mathbf{Y}}\!\in\!(0,1]^{N\times C}$ the model probabilities (softmax of the logits). The loss is
\begin{equation}
\label{eq:balanced_ce}
\mathcal{L}_{\mathrm{BaCE}}(\mathbf{Y},\hat{\mathbf{Y}})
\;=\; -\,\frac{1}{N}\sum_{i=1}^{N}\sum_{c=1}^{C}\beta_c\,Y_{ic}\,\log\hat{Y}_{ic}\,,
\end{equation}
where $\beta_c>0$ are class-balance coefficients. Intuitively, Eq.~\ref{eq:balanced_ce} increases the penalty for underrepresented classes (e.g., \texttt{QSO\_high}) so the loss is not dominated by more abundant types such as \texttt{STAR}/\texttt{GALAXY}. We compute the weights $\beta_c$ from the empirical frequencies $\hat{\pi}_c$ of the training splits via inverse-frequency normalization:
\begin{equation}
\label{eq:beta_weights}
\beta_c \;=\; \frac{\hat{\pi}_c^{-1}}{\frac{1}{C}\sum_{k=1}^{C}\hat{\pi}_k^{-1}}.
\end{equation}
During training, we also include standard weight regularization ($\ell_2$ term in the loss) to improve generalization and stabilize optimization.

\subsection{Models and training regimes}\label{subsec:regimes}

All experiments share the input representation and preprocessing from Sect.~\ref{subsec:datasets_curation}: 55 J-PAS bands after reliability masking concatenated with \textit{Tractor} morphology flags. The different classifiers we are going to train are comprised of an encoder $f_{\theta}$ and a downstream head $g_{\phi}$ (both multi-layer perceptrons) producing a four-way softmax over \texttt{QSO\_high}, \texttt{QSO\_low}, \texttt{GALAXY}, \texttt{STAR}. We compare three regimes:

\paragraph{(i) Mocks-only (no-DA) zero-shot transfer.}
Train $(f_\theta,g_\phi)$ on DESI$\rightarrow$J-PAS \emph{mocks} using Eq.~\ref{eq:balanced_ce}; select the checkpoint that minimizes the \emph{mock-validation} BaCE (In practice, ranking by mock-validation BaCE or by mock-validation macro-F1 leads to the same top-performing models). This baseline measures the raw sim$\rightarrow$obs gap once we test the model on real \emph{J-PAS observations}.

\paragraph{(ii) Semi-supervised domain adaptation (SSDA; DA).}
Start from several top-performing no-DA checkpoints, \emph{freeze the downstream head} $g_\phi$ to keep the decision boundaries (learned from abundant source labels) fixed, and \emph{re-train only the encoder} $f_\theta$ on the labeled J-PAS \emph{training} split so that target-domain features align with the head’s fixed class regions. Concretely, during SSDA we keep all head downstream model parameters fixed, while the encoder remains trainable. This step warps features so that J-PAS examples fall into the correct class-conditioned regions while keeping the decision surfaces fixed, improving alignment and calibration with modest target supervision.

\paragraph{(iii) J-PAS-supervised (target-only) baseline.}
Train $(f_\theta,g_\phi)$ from scratch on the labeled J-PAS \emph{training} split and select by the minimum \emph{J-PAS-validation} BaCE. This baseline quantifies performance when relying exclusively on target labels.

\subsection{Hyperparameter search and convergence}\label{subsec:hpo}

To ensure convergence and a fair comparison between models, we ran a hyperparameter sweep using the \texttt{Weights \& Biases} package~\citep{wandb} for each regime over network depth/width, dropout, learning rate and weight decay, batch size, and gradient clipping. Model selection uses the validation loss appropriate to each regime: mock-validation BaCE for the no-DA pretraining; J-PAS-validation BaCE for SSDA and J-PAS-supervised models. The top configurations cluster tightly in validation metrics, and re-runs with different seeds reproduce the same training plateaus, indicating that results are not driven by lucky initializations or idiosyncratic settings. 

All configuration files (\texttt{wandb} sweep specs, seeds, and exact hyperparameters) and the best trained checkpoints for the three regimes are publicly available at:
\url{https://github.com/daniellopezcano/JPAS_Domain_Adaptation}

\section{Results}\label{sec:results}

\begin{figure*}
  \centering
  \includegraphics[width=\textwidth]{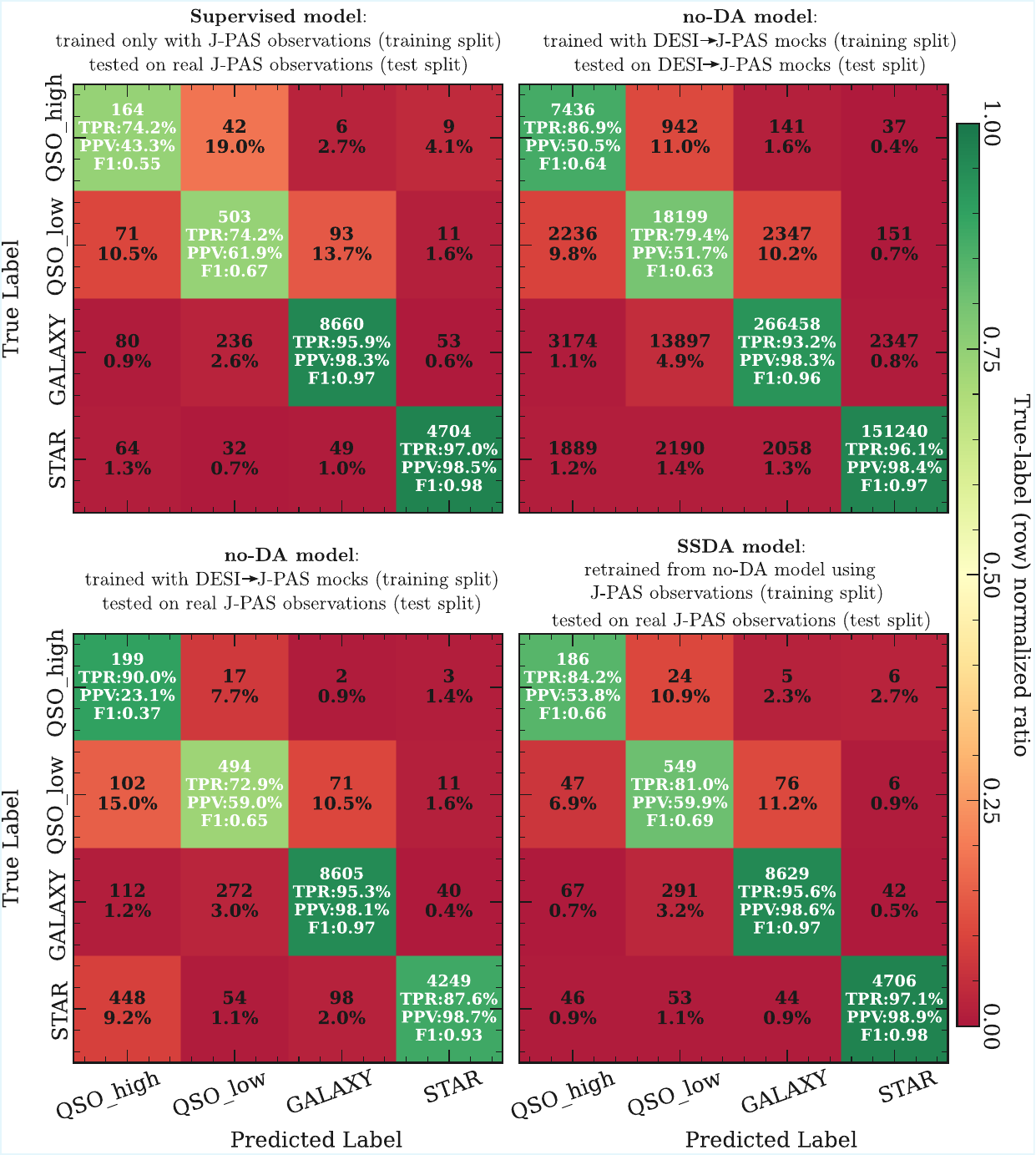}
  \caption{
  Confusion matrices for four complementary evaluations:
  (top left) \emph{J-PAS supervised}: target-only baseline,
  (top right) \emph{Mocks no-DA}: source-domain evaluation on the mock test split,
  (bottom left) \emph{J-PAS no-DA}: zero-shot transfer of the mocks-trained model,
  and (bottom right) \emph{J-PAS DA}: the same model after semi-supervised adaptation.
  Cells show \emph{counts}; the colormap is \emph{row-normalized} to aid visual comparison. Diagonals also list TPR (recall), PPV (precision), and F1 (see Eq.~\ref{eq:tpr_ppv_f1}) for each class.
  }
  \label{fig:confusion-matrices}
\end{figure*}

All results in this section use held-out test splits that were never employed during the training/validation process (see Sect.~\ref{sec:methods}). On the \emph{target} side, we evaluate on the reserved J-PAS Observations test split with an evaluation-only cut of $r \le 22.5$ (see Sect.~\ref{sec:data}). On the \emph{source} side, we evaluate the results on the DESI$\rightarrow$J-PAS \emph{mock} test split to establish in-domain performance of the \emph{mocks-only} model and quantify the sim$\to$obs gap.
After our selection cuts, the J-PAS test split contains $N_{\mathrm{J\text{-}PAS}}^{\mathrm{test}}=14{,}777$ objects, and the corresponding mocks test split contains $\approx 10^6$ individual sources.

We report the \emph{best} model in each regime -- J-PAS-supervised, Mocks no-DA, and DA -- selected using the validation criteria defined in Sect.~\ref{subsec:regimes}.

Figure~\ref{fig:confusion-matrices} compiles four confusion matrices\footnote{
A confusion matrix is a cross-tabulation of \emph{true} (rows) versus \emph{predicted} (columns) class counts; diagonal cells are correct classifications, and off-diagonals quantify specific misclassifications. 
Row-normalization makes the color scale comparable across rows/classes.
}, one per training/evaluation regime. (Top-left) \emph{J-PAS supervised}: target-only training, evaluated on J-PAS observations test split, (Top-right) \emph{Mocks no-DA (in-domain)}: trained and evaluated on the DESI$\rightarrow$J-PAS mock splits, (Bottom-left) \emph{J-PAS no-DA (zero-shot)}: trained on mocks, evaluated on J-PAS test observations, and (Bottom-right) \emph{J-PAS DA (SSDA)}: mocks-pretrained, encoder adapted with a small labeled J-PAS set while keeping the head fixed and then evaluated on the J-PAS observations test split. Each cell shows the absolute number of objects in that predicted/true bin, while the colormap is row-normalized to the total number of sources in the corresponding true class to aid cross-class comparison. For the diagonal entries we additionally report three per-class diagnostics: the true positive rate (TPR/recall), the positive predictive value (PPV/precision), and the F1 score, defined from the usual confusion-matrix counts (true positives TP, false positives FP, false negatives FN) as
\begin{equation}
\label{eq:tpr_ppv_f1}
\begin{aligned}
    \mathrm{TPR}&=\frac{\mathrm{TP}}{\mathrm{TP}+\mathrm{FN}},\\
    \mathrm{PPV}&=\frac{\mathrm{TP}}{\mathrm{TP}+\mathrm{FP}},\\
    \mathrm{F1}&=\frac{2\,\cdot \mathrm{TPR}\cdot\,\mathrm{PPV}}{\mathrm{TPR}+\mathrm{PPV}}.
\end{aligned}
\end{equation}

Across all panels, the residual off--diagonal counts follow physically expected effects (see Appendix~\ref{app:qso-probs-vs-z}, Fig.~\ref{fig:P_vs_z_QSOs}). 
First, at low redshift (\(z\!\lesssim\!0.3\)) host-galaxy light dilutes AGN features in the J-PAS narrow bands, shifting probability from \texttt{QSO\_low} toward \texttt{GALAXY}. 
Second, near \(z\simeq2.1\), the separation between \texttt{QSO\_low} and \texttt{QSO\_high} is set by an explicit redshift threshold, so any method that effectively infers \(z\) from \emph{J-spectra} (either explicitly or implicitly, as in our case) will show unavoidable uncertainty-driven mixing across the boundary. Finite filter width and measurement noise limit the redshift precision, and quasars with true redshifts near $2.1$ naturally receive similar support for both subclasses.

Third, for \emph{true} \texttt{QSO\_high} at higher redshift, narrow redshift intervals show additional leakage into \texttt{QSO\_low} consistent with line-identification aliases---e.g., configurations where Ly$\alpha$ or C\,\textsc{iv} falling in specific bands can mimic the lower-\(z\) template dominated by Mg\,\textsc{ii} in the J-PAS passbands. 
Appendix~\ref{app:qso-probs-vs-z} quantifies these behaviours with probability--vs--\(z\) medians and dispersions, including a zoom that confirms the structural ambiguity at \(z\simeq2.1\).

Differences between panels reflect the interplay of dataset sizes and domain shift. The \emph{Mocks no-DA} matrix (tested employing in-domain mock observations; top-right) shows very tight behaviour, consistent with training and evaluation sharing the same synthetic distribution. However, when the same model is applied zero-shot to J-PAS (bottom-left), the off-diagonal leakage increases, exposing the sim$\rightarrow$obs shift discussed in Sect.~\ref{sec:intro}. A clear effect in the \emph{J-PAS no-DA (zero-shot)} panel appears as extra confusion between \texttt{STAR} and QSO classes: in mocks about $1.2\%$ of \texttt{STAR} are labelled as \texttt{QSO\_high}, while on real J-PAS observations this rises to $9.2\%$.
After SSDA, the confusion drops to $\simeq 0.9\%$, comparable to the in-domain mock value, indicating that the $\sim$8\% excess in the zero-shot case is predominantly a sim$\rightarrow$obs domain-shift effect rather than a physical degeneracy.

The \emph{J-PAS supervised} baseline (top-left) is competitive overall but constrained by the limited training dataset and imbalanced target-label budget. The rare-class boundaries are less stable, and higher levels of \texttt{QSO\_high}/\texttt{QSO\_low} confusion and \texttt{QSO\_low}/\texttt{GALAXY} mixing remains. By contrast, the \emph{J-PAS DA} panel (bottom-right) shows a visible re-concentration of counts along the diagonal for both quasar classes, while \texttt{GALAXY}/\texttt{STAR} remain essentially saturated. In other words, the adaptation step warps the encoder so that J-PAS examples fall into the correct class-conditional regions constructed from abundant source labels (Sect.~\ref{subsec:regimes}), reducing the ambiguities expected from spectral physics and measurement noise. Visual inspection already suggests that SSDA yields the best on-target performance; to avoid qualitative comparisons, we now turn to compact, class-resolved metrics. A per-class breakdown of accuracy, F1, recall, precision, `area-under-curve' (AUC), and calibration is provided in Appendix~\ref{app:diagnostics}.

\begin{figure}
  \centering
  \includegraphics[width=1.0\columnwidth]{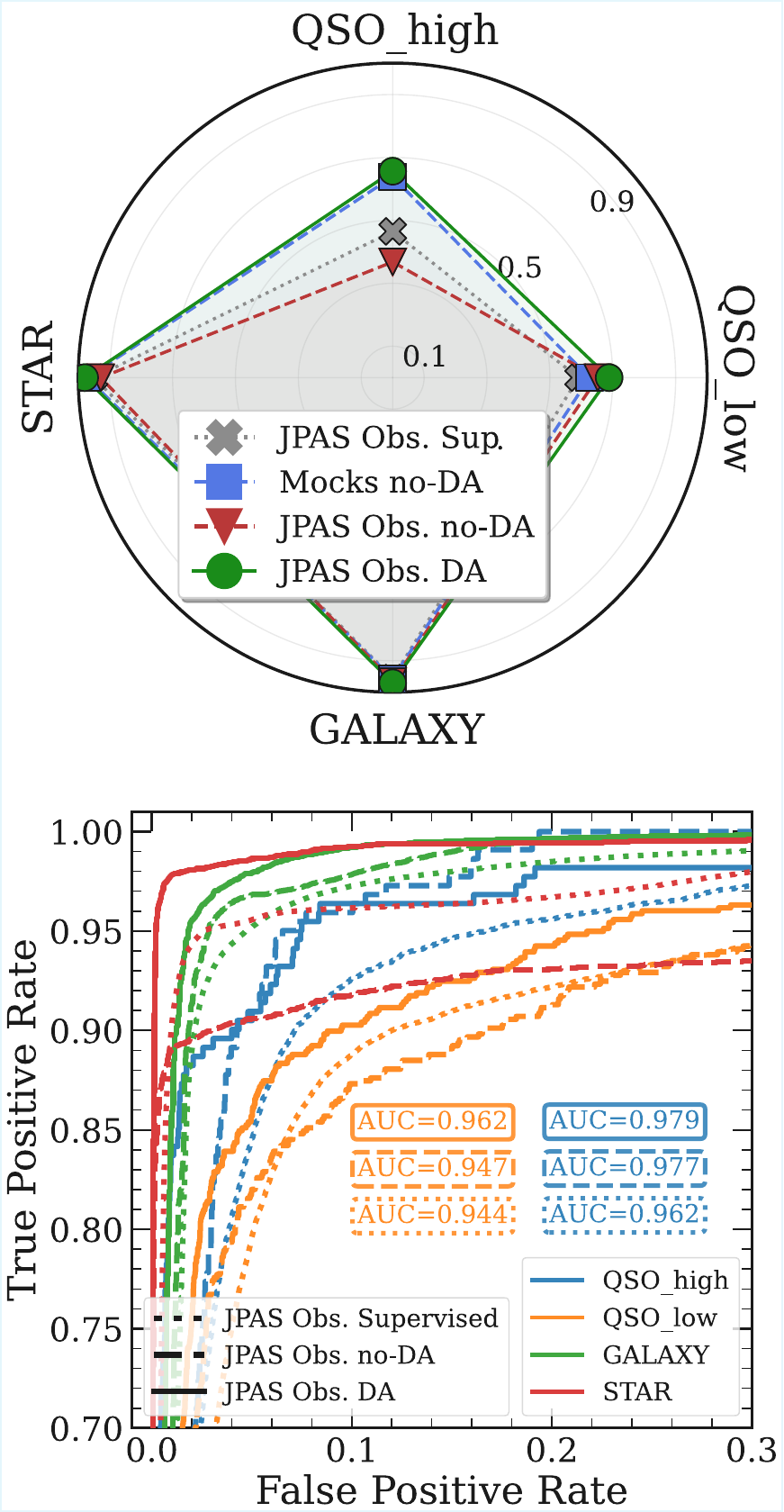}
  \caption{
  \textbf{Per-class summary diagnostics.}
  \emph{Top:} F1 scores by class (radar plot) shown for all four regimes (different markers with distinct colors and linestyles).
  \emph{Bottom:} Per-class ROC curves and AUC values; the low-FPR region (left) is most relevant for building clean follow-up target lists. 
  Both panels reflect the same trend seen in Fig.~\ref{fig:confusion-matrices}: SSDA boosts the quasar subclasses while leaving \texttt{GALAXY}/\texttt{STAR} essentially saturated.
  }
  \label{fig:F1_and_ROC_vert}
\end{figure}

Figure~\ref{fig:F1_and_ROC_vert} summarizes per-class behavior in two complementary ways. The top panel (radar plot) shows the F1-scores by class; this makes the rare-class trade-offs easy to read. The global performance of the models can be interpreted as the enclosed area defined by the different closed-curves. The bottom panel displays per-class ROC curves on test-split J-PAS observations with AUC values. The low-FPR region (left side) is particularly relevant for constructing clean candidate lists for follow-up surveys. Both views tell a consistent story: SSDA lifts the performance for the two QSO subclasses simultaneously: F1 increases and the Receiver Operating Characteristic (ROC) curves shift up/left---while leaving \texttt{GALAXY}/\texttt{STAR} essentially unchanged. Minor dips for a majority class can occur (the training objective is the balanced cross-entropy; Sect.~\ref{subsec:balanced_ce}), but the net effect is positive: macro metrics improve and the label-scarce, shift-sensitive QSO classes benefit the most.
SSDA is optimized to improve target-domain performance under our training objective loss (Eq.~\ref{eq:balanced_ce}), but it does not guarantee strictly monotonic gains in every derived local metric; for \texttt{QSO\_high}, for instance, the improvement of the $F_1$-score and the PPV value can come with a reduction ($\sim 6\%$, see Figure~\ref{fig:confusion-matrices}) in the TPR value.

To guide operational thresholds for future spectroscopic surveys focused on QSO detection such as WEAVE-QSO, Appendix~\ref{app:diagnostics} examines how class-wise F1 varies with simple $r$-band magnitude cuts, see Fig.~\ref{fig:magnitudes_F1_radar}.

Table~\ref{tab:metrics_four_cases} consolidates global statistics on the corresponding test-splits for the four different experiments. We report overall accuracy ($\mathrm{Acc}$), macro-averaged F1, TPR/recall, PPV/precision, AUC, and the Expected Calibration Error (ECE), a standard summary statistic of probabilistic calibration computed by binning predictions by confidence. With $M$ bins $B_m$ of sizes $n_m$, empirical accuracies $\mathrm{acc}(B_m)$, and mean confidences $\mathrm{conf}(B_m)$,
\begin{equation}
\label{eq:ece}
\mathrm{ECE}=\sum_{m=1}^{M}\frac{n_m}{N}\,\big|\mathrm{acc}(B_m)-\mathrm{conf}(B_m)\big|\,.
\end{equation}
Higher values indicate better performance except for ECE (where lower means better). ``Macro'' scores are unweighted means of the per-class values. The AUC summarizes global threshold-free ranking quality. 

As anticipated from the confusion-matrix trends, the SSDA model delivers the strongest \emph{on-target} performance on J-PAS across most metrics: it attains the highest accuracy (0.952), macro-F1 (0.824), macro-TPR/recall (0.894), and macro-PPV/precision (0.778).
For the AUC score, the \emph{Mocks no-DA (in-domain)} model reaches 0.977---unsurprising, since this evaluates ranking on the same synthetic distribution used for training. On real J-PAS test-set observations, SSDA's AUC (0.975) is the best among the J-PAS evaluations and consistent with its gains elsewhere. Regarding calibration, the lowest ECE is observed for \emph{J-PAS no-DA} model (0.045) while SSDA is close (0.048); these close values indicate similarly good calibration in practice. The \emph{J-PAS supervised} baseline remains competitive but is ultimately limited by the scarcity and class imbalance of labelled \texttt{QSO\_high}, which SSDA mitigates by leveraging abundant source labels plus modest target supervision. These aggregate numbers are consistent with the class-resolved trends summarized in Appendix~\ref{app:diagnostics}.

\begin{table}
\caption{Global metrics on the test splits for each experiment. Column keys correspond to: J-sup (J-PAS supervised), M-noDA (mocks in-domain), J-noDA (zero-shot from mocks to J-PAS), and J-SSDA (domain-adapted). We summarize overall accuracy (Acc), macro-F1, macro recall (TPR), macro precision (PPV), macro area under the ROC curve (AUC), and expected calibration error (ECE; lower is better).}
\label{tab:metrics_four_cases}
\centering
{\small\setlength{\tabcolsep}{4pt}
\begin{tabular}{ccccc}
\hline\hline
Metric & J-sup & M-noDA & J-noDA & J-SSDA \\
\hline
Acc  & 0.950 & 0.934 & 0.917 & \textbf{0.952} \\
F1   & 0.792 & 0.798 & 0.729 & \textbf{0.824} \\
TPR  & 0.853 & 0.889 & 0.865 & \textbf{0.894} \\
PPV  & 0.755 & 0.747 & 0.697 & \textbf{0.778} \\
AUC  & 0.957 & \textbf{0.977} & 0.959 & 0.975 \\
ECE  & 0.054 & 0.047 & \textbf{0.045} & 0.048 \\
\hline
\end{tabular}}
\end{table}

\section{Conclusions}\label{sec:conclusions}

Domain shift is transversal across many studies in astronomy and cosmology: simulators and overlapping surveys do not share identical characteristics, so the performance of most models drops when deployed out of domain. Here we focused on a concrete case, J-PAS four-way source classification, and asked how far a minimal Semi-Supervised Domain Adaptation step can bridge the simulations-to-observations gap.

The approach is simple. We pretrain on DESI$\rightarrow$J-PAS \emph{mocks} (DESI spectra convolved with the J-PAS narrow-band transmission curves to yield \emph{J-spectra}-like with DESI labels) and then nudge the representation to real J-PAS by adapting only the encoder on a small, leak-safe labeled subset while keeping the classification head fixed. On held-out J-PAS data, our adapted model improves accuracy to $0.952$ (from $0.917$), macro-F1 to $0.824$ (from $0.729$), and AUC to $0.975$ (from $0.959$), with well-behaved calibration ($\mathrm{ECE}\!\approx\!0.05$). The quasar subclasses benefit most: the distinction between high and low redshift quasars (split at $z = 2.1$) is cleaner and confusion with galactic sources is reduced; for example, $\mathrm{F1}(\texttt{QSO\_high})$ reaches $0.66$ versus $0.55$ (J-PAS--supervised) and $0.37$ (no-DA zero-shot).

For WEAVE-QSO survey~\citep{2016sf2a.conf..259P, 2024MNRAS.530.2688J}, calibrated class probabilities, especially for quasars, support clean candidate lists built at low false-positive rates; PPV-based yield forecasts allow field-by-field budgeting. The adaptation is data-efficient, so periodic refreshes can track slow changes in seeing or zeropoints. Keeping per-object selection probabilities makes the selection function explicit for $n(z)$ and clustering analyses.

Looking ahead, the same techniques applied in this work can be extended to related studies involving cross-survey classification, photo-$z$ estimation, and forward-modeling pipelines. Contrastive/self-supervised pretraining on unlabeled data can learn domain-agnostic embeddings that transfer well across different simulation-specific characteristics (sub-grid physics, resolution effects, etc.) and instrument specifications (passbands, depths, and PSFs); coupled with domain adaptation techniques, this helps bridge survey-to-survey and sim-to-obs gaps by reducing bias and calibration drift, preserving class boundaries, and improving coverage and likelihood calibration. We encourage treating such domain-aware methods as standard practice so that large simulated or auxiliary datasets translate into reliable, reproducible results on the target observations.


\begin{acknowledgments}
We are grateful to Rodrigo Voivodic and Natalia Villa Nova Rodrigues for insightful discussions, constructive suggestions, and many helpful comments throughout this work.

DLC is supported by the São Paulo Research Foundation (FAPESP) via a Postdoctoral Fellowship (Process No.~2024/05768-9), linked to the FAPESP Thematic Project (Process No.~2019/26492-3). 

RA acknowledges support from the São Paulo Research Foundation (FAPESP) through the same Thematic Project (Process No.~2019/26492-3) and from the Brazil--France FAPESP--ANR program (FAPESP Process No.~2022/03426-8).

IPR was supported by funding from the grant PID2023-151122NA-I00 by MICIU/AEI/10.13039/501100011033 and by ERDF/EU.

G.M.S. acknowledges financial support from Severo Ochoa grant CEX2021-001131S funded by MICIU/AEI/\mbox{10.13039/501100011033} and project PID2022-141755NB-I00. G.M.S. also acknowledges support from AST22\_00001\_Subp 26 and 11, funded by the European Union--NextGenerationEU; the Ministerio de Ciencia, Innovación y Universidades; the Plan de Recuperación, Transformación y Resiliencia; the Consejería de Universidad, Investigación e Innovación (Junta de Andalucía); and the Consejo Superior de Investigaciones Científicas

RGD acknowledges financial support from the project PID2022-141755NB-I00 and the Severo Ochoa grant CEX2021-001131-S, funded by MICIU/AEI/10.13039/501100011033.

This research used data obtained with the Dark Energy Spectroscopic Instrument (DESI). DESI construction and operations is managed by the Lawrence Berkeley National Laboratory. This material is based upon work supported by the U.S. Department of Energy, Office of Science, Office of High-Energy Physics, under Contract No. DE--AC02--05CH11231, and by the National Energy Research Scientific Computing Center, a DOE Office of Science User Facility under the same contract. Additional support for DESI was provided by the U.S. National Science Foundation (NSF), Division of Astronomical Sciences under Contract No. AST-0950945 to the NSF's National Optical-Infrared Astronomy Research Laboratory; the Science and Technology Facilities Council of the United Kingdom; the Gordon and Betty Moore Foundation; the Heising-Simons Foundation; the French Alternative Energies and Atomic Energy Commission (CEA); the National Council of Humanities, Science and Technology of Mexico (CONAHCYT); the Ministry of Science and Innovation of Spain (MICINN), and by the DESI Member Institutions: \url{https://www.desi.lbl.gov/collaborating-institutions}. The DESI collaboration is honored to be permitted to conduct scientific research on I'oligam Du'ag (Kitt Peak), a mountain with particular significance to the Tohono O'odham Nation. Any opinions, findings, and conclusions or recommendations expressed in this material are those of the author(s) and do not necessarily reflect the views of the U.S. National Science Foundation, the U.S. Department of Energy, or any of the listed funding agencies.

This paper has gone through an internal review by the J-PAS collaboration. Based on observations made with the JST/T250 telescope and JPCam at the Observatorio Astrofísico de Javalambre (OAJ), in Teruel, owned, managed, and operated by the Centro de Estudios de Física del Cosmos de Aragón (CEFCA). We acknowledge the OAJ Data Processing and Archiving Unit (UPAD) for reducing and calibrating the OAJ data used in this work. Funding for the J-PAS Project has been provided by the Governments of Spain and Arag\'on through the Fondo de Inversiones de Teruel; the Aragonese Government through the Research Groups E96, E103, E16\_17R, E16\_20R, and E16\_23R; the Spanish Ministry of Science and Innovation (MCIN/AEI/10.13039/501100011033 y FEDER, Una manera de hacer Europa) with grants PID2021-124918NB-C41, PID2021-124918NB-C42, PID2021-124918NA-C43, and PID2021-124918NB-C44; the Spanish Ministry of Science, Innovation and Universities (MCIU/AEI/FEDER, UE) with grants PGC2018-097585-B-C21 and PGC2018-097585-B-C22; the Spanish Ministry of Economy and Competitiveness (MINECO) under AYA2015-66211-C2-1-P, AYA2015-66211-C2-2, and AYA2012-30789; and European FEDER funding (FCDD10-4E-867, FCDD13-4E-2685). The Brazilian agencies FINEP, FAPESP, FAPERJ and the National Observatory of Brazil have also contributed to this project. Additional funding was provided by the Tartu Observatory and by the J-PAS Chinese Astronomical Consortium.

\end{acknowledgments}

\begin{contribution}


Contributions are described using the CRediT (Contributor Roles Taxonomy; ANSI/NISO) framework (\url{https://credit.niso.org}).

\begin{description}
\item[\textbf{DLC}] \emph{Conceptualization}: defined the sim$\rightarrow$obs transfer problem and the domain adaptation pipeline. \emph{Methodology}: designed the SSDA recipe, loss function, and evaluation metrics. \emph{Software}: implemented data pipeline, training/evaluation code, and \texttt{wandb} sweeps. \emph{Data curation}: created leak-safe splits; applied reliability masking; standardized bands. \emph{Formal analysis}: Trained the different classification models; provided physical interpretation of the results. \emph{Validation}: computed evaluation metrics (macro-F1, TPR/PPV, AUC, ECE). \emph{Visualization}: produced all figures. \emph{Investigation}: carried out a literature review of previous works employing DA/CL techniques in astrophysics and cosmology. \emph{Resources}: packaged the public repository; built the machine where the analysis was performed. \emph{Writing\textemdash original draft}: wrote the manuscript core sections and appendix materials.

\item[\textbf{RA}] \emph{Conceptualization}: helped frame the sim$\rightarrow$obs strategy and target metrics. Contributed to scientific discussions within the QSO identification working group. \emph{Resources}: generated DESI$\rightarrow$J\textendash PAS mock catalogs and the instrument/noise model used to render \emph{J-spectra}. \emph{Writing\textemdash review \& editing}: provided iterative feedback on methods, results, and interpretation. \emph{Funding acquisition}: secured support via the FAPESP Thematic Project (Process No.~2019/26492\textendash3) and the Brazil--France FAPESP--ANR program (FAPESP Process No.~2022/03426\textendash8), enabling data preparation, compute resources, and personnel time.

\item[\textbf{LN}] \emph{Data curation}: cleaned J\textendash PAS IDR inputs, applied reliability/coverage masks. \emph{Resources}: produced the cross-matched catalogs and raw data products. \emph{Writing\textemdash review \& editing}: verified data descriptions and clarified pipeline details. \emph{Conceptualization}: contributed to scientific discussions within the QSO identification working group.

\item[\textbf{IPR}] \emph{Data curation}: performed the J\textendash PAS$\times$DESI cross-matching and prepared the final match tables. \emph{Resources}: constructed the raw datasets delivered to the analysis pipeline. \emph{Writing\textemdash review \& editing}: reviewed the data construction and results sections. \emph{Conceptualization}: contributed to scientific discussions within the QSO identification working group.

\item[\textbf{GMS}] \emph{Writing---review \& editing}: provided detailed feedback on multiple drafts of the manuscript, particularly regarding quasar classification strategy. \emph{Conceptualization}: contributed to scientific discussions within the QSO identification working group.

\item[\textbf{JCM}] \emph{Writing\textemdash review \& editing}: reviewed the final manuscript and provided useful suggestions to this work.

\end{description}

All authors reviewed and approved the final manuscript.
\end{contribution}

%
\facilities{
We use observations from the J-PAS survey obtained with JPCam at the Observatorio Astrofísico de Javalambre (OAJ).
}

\software{
\textsc{Python},
\textsc{Matplotlib} \citep{Matplotlib-Hunter07},
\textsc{PyTorch} \citep{2019arXiv191201703P},
\textsc{Numpy} \citep{Numpy-vanDerWalt11},
\textsc{Weights \& Biases} \citep{wandb},
Source Extractor \citep{1996A&AS..117..393B}
}


\appendix

\section{Quasar class probabilities as a function of redshift}\label{app:qso-probs-vs-z}

All diagnostics in this Appendix are computed using the \emph{no-DA} model evaluated on the DESI$\!\rightarrow$J-PAS \emph{mocks}.
We use the mock sample here because it provides much larger statistics, making broad trends and fine structure in the probability--vs--$z$ behaviour easier to visualize.

\begin{figure*}
\plotone{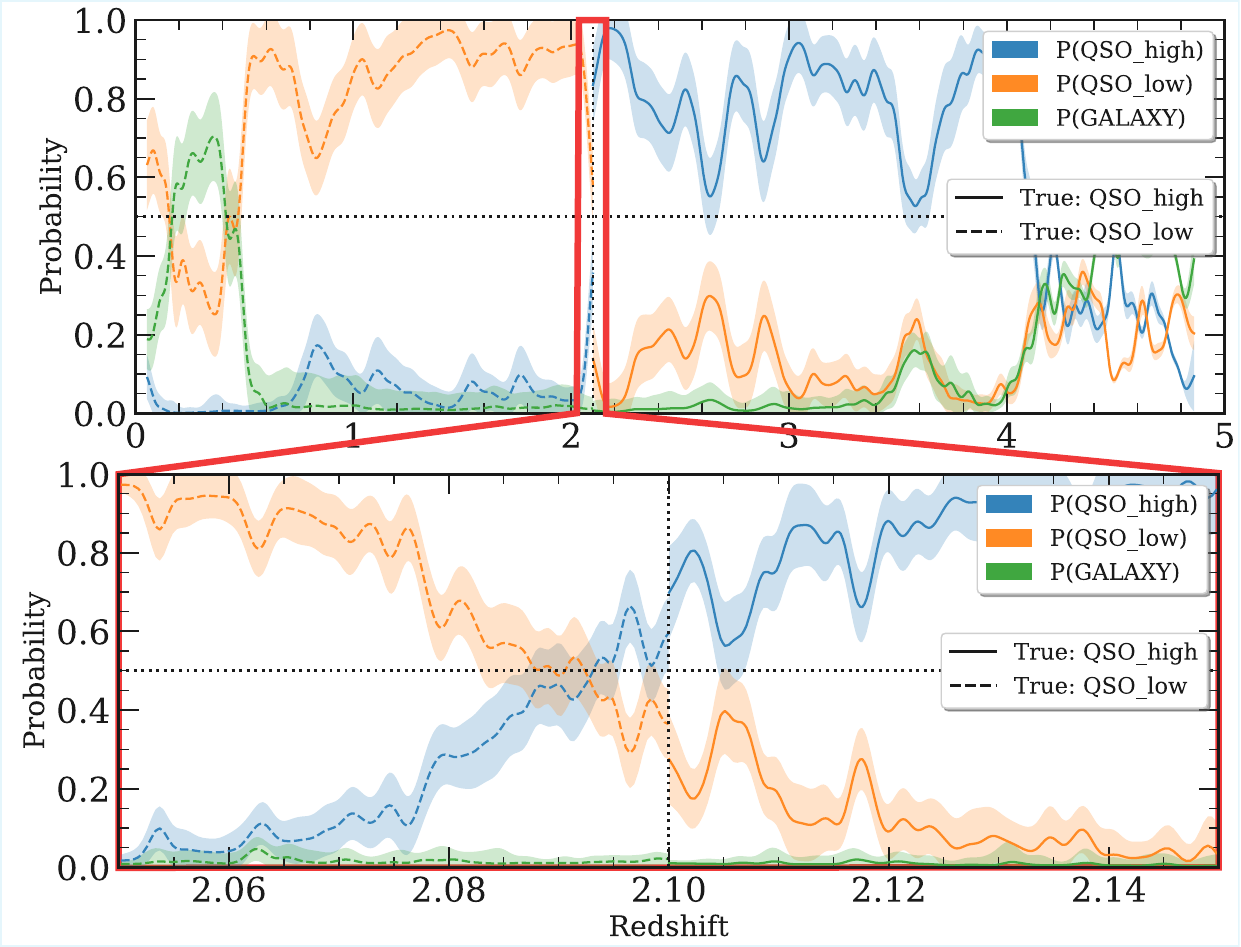}
\caption{
\textbf{Quasar class probabilities as a function of redshift (two-panel view).}
\emph{Top:} range \(0<z<5\).
\emph{Bottom:} zoom on the transition near \(z\simeq2.1\) (red box in the top panel marks the window magnified below).
For objects whose \emph{true} labels are \texttt{QSO\_high} (solid curves) or \texttt{QSO\_low} (dashed curves), we plot the predicted class probabilities as a function of spectroscopic redshift:
blue = \(P(\texttt{QSO\_high})\),
orange = \(P(\texttt{QSO\_low})\),
green = \(P(\texttt{GALAXY})\) (the \texttt{STAR} channel is omitted for clarity).
Curves show the \emph{binned median} probability in redshift,  smoothed for readability;
shaded bands indicate a conservative dispersion of \(\pm\frac{1}{3}\sigma\) (one third of the per-bin sample standard deviation) around each median.
The vertical dotted line denotes the subclass split at \(z=2.1\); the horizontal dotted line marks the reference level \(p=0.5\).
Together, the panels visualize the two main sources of ambiguity discussed in the text: low-\(z\) host dilution (\texttt{QSO\_low} vs.\ \texttt{GALAXY}) and mutual \texttt{QSO\_high}/\texttt{QSO\_low} confusion at the \(z\simeq 2.1\) boundary.
}
\label{fig:P_vs_z_QSOs}
\end{figure*}

To interpret the off--diagonal structure of the confusion matrices presented in section~\ref{sec:results}, we examine how the predicted class probabilities vary with redshift for objects whose \emph{true} labels are \texttt{QSO\_high} or \texttt{QSO\_low}. Figure~\ref{fig:P_vs_z_QSOs} shows two views of the same diagnostics: the \textbf{top panel} covers the redshift range (\(0<z<5\)), and the \textbf{bottom panel} zooms into the transition near \(z\simeq 2.1\).

Each curve is the binned \emph{median} probability as a function of redshift; shaded bands indicate a conservative dispersion of \(\pm \frac{1}{3}\sigma\) around the median.

The vertical dotted line marks \(z=2.1\) (our \texttt{QSO\_high}/\texttt{QSO\_low} split) and the horizontal line marks \(p=0.5\). Solid and dashed linestyles indicate the \emph{true} class label (\texttt{QSO\_high} vs.\ \texttt{QSO\_low}).

\paragraph{Top panel: global trends (\(0<z<5\)).}
Two regimes explain most of the \texttt{QSO\_low} confusion:
\begin{enumerate}
    \item At low redshift (\(z\!\sim\!0.0\)--\(0.5\)), the median \(p(\texttt{GALAXY})\) rises while \(p(\texttt{QSO\_low})\) softens. This matches the expectation that host-galaxy light increasingly dilutes the AGN signal in the J-PAS narrow bands,
    the \emph{J-spectra} of low-\(z\) AGN can be dominated by resolved host-galaxy emission when the nucleus is comparatively faint, and the classification may be driven by host-galaxy properties rather than by nuclear emission alone.

    \item Near the subclass boundary (\(z\simeq2.1\)), \(p(\texttt{QSO\_low})\) and \(p(\texttt{QSO\_high})\) converge, {as expected for a hard redshift threshold combined with finite \emph{J-spectra} resolution: finite filter widths and measurement noise limit the effective redshift precision, so objects with true redshifts near $2.1$ naturally receive comparable support for both subclasses.}    
\end{enumerate}

For \texttt{QSO\_high}, the same boundary effect appears around \(z\simeq2.1\). At higher redshift the curves show additional, narrower bumps where \(p(\texttt{QSO\_high})\) dips and \(p(\texttt{QSO\_low})\) momentarily increases. These features are consistent with occasional line-identification aliases---effectively, the classifier infers redshift from spectral features, and certain filter/line coincidences can transiently favour the wrong subclass. At the highest redshifts there are very few detected sources, and the smaller S/N for these objects results in uncertain class probabilities.

\paragraph{Bottom panel: zoom on the transition (\(2.05\!\lesssim\!z\!\lesssim\!2.15\)).}
The crossover is clearer in the zoom: the median \(p(\texttt{QSO\_high})\) climbs through \(\sim 0.5\) as \(z\) crosses $2.1$ while \(p(\texttt{QSO\_low})\) symmetrically falls, for both true-label subsets. The overlap of the dispersion bands confirms that the ambiguity at the split is \emph{structural}---set by J-PAS filter sampling of QSO features---rather than a mere modelling artifact.

In the smoothed-binned curves, the apparent $p(\texttt{QSO\_high})\!=\!p(\texttt{QSO\_low})$ crossing can occur slightly offset from the nominal threshold (e.g., by $\Delta z\sim 0.01$). We do not assign a physical meaning to such a small displacement: it can arise from finite sample size and the binning/smoothing procedure used to summarize the point cloud in redshift, and it can also reflect a purely \emph{model-side} effect---i.e., a particular converged solution that yields essentially the same global objective value but induces a small local shift in the median curves near the boundary, without implying an underlying feature in the data. Because this diagnostic is meant to be illustrative (and because resolving the detailed origin of percent-level offsets would require a dedicated stability/ablation analysis beyond the scope of this Appendix), we caution against over-interpreting the exact crossing location.

{For completeness, we reiterate that Fig.~\ref{fig:P_vs_z_QSOs} uses the no-DA model on mocks to maximize statistical power; the same qualitative behaviours are present in the J-PAS observations, but the smaller sample size makes the trends noisier.}

\section{Additional per-class diagnostics and magnitude-cut trade-offs}\label{app:diagnostics}

\begin{figure*}
\centering
\includegraphics[width=0.7\textwidth]{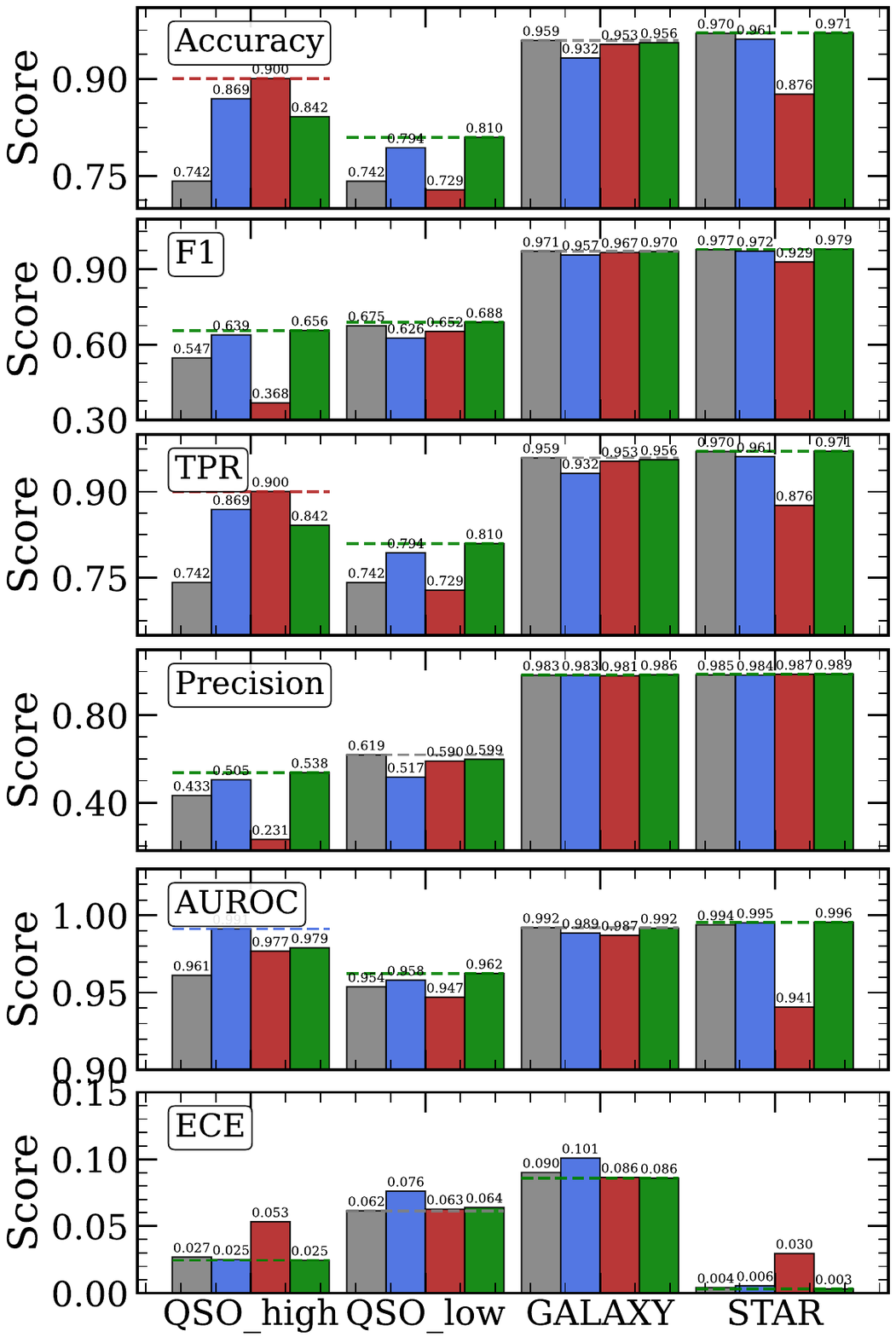}
\caption{
\textbf{Per--class diagnostics across regimes.}
Panels show, from top to bottom: \emph{Accuracy}, \emph{F1}, \emph{TPR/Recall}, \emph{Precision/PPV}, \emph{AUC}, and \emph{ECE}.
Within each panel, bars are grouped by true class (\texttt{QSO\_high}, \texttt{QSO\_low}, \texttt{GALAXY}, \texttt{STAR}); values are printed above the bars and dashed horizontal lines mark the per--class top-performance model.
Colors are fixed across all panels and figures: 
\textbf{gray} = \emph{J-PAS supervised} (evaluated on the J-PAS Observations test split), 
\textbf{blue} = \emph{Mocks no-DA in-domain} (trained and evaluated on DESI$\rightarrow$J-PAS mocks), 
\textbf{red} = \emph{J-PAS no-DA} (trained on mocks, evaluated zero-shot on the J-PAS Observations test split), and 
\textbf{green} = \emph{J-PAS SSDA} (mocks-pretrained, encoder adapted on a small labeled J-PAS subset; evaluated on the J-PAS Observations test split).
}
\label{fig:per-class-metrics}
\end{figure*}

Figure~\ref{fig:per-class-metrics} collects per-class metrics across the four evaluation regimes discussed in Sect.~\ref{sec:methods} and~\ref{sec:results}. Each panel reports one diagnostic (accuracy, F1, TPR/recall, precision/PPV, AUC, and ECE), with bars grouped by class (\texttt{QSO\_high}, \texttt{QSO\_low}, \texttt{GALAXY}, \texttt{STAR}). The plot makes explicit where adaptation helps most and where performance is already saturated. For the quasar subclasses, SSDA tightens recall and precision together, lifting F1 while keeping calibration at the $\sim$\,few-percent level; the \texttt{STAR} and \texttt{GALAXY} metrics remain high and flat across regimes, consistent with the confusion matrices in Fig.~\ref{fig:confusion-matrices}. Dashed horizontal guides indicate reference levels within each panel and help compare the regimes class-by-class without relying on a single macro average. Read together, these panels support the main-text claim that adaptation reduces star$\rightarrow$QSO leakage and sharpens the \texttt{QSO\_high}/\texttt{QSO\_low} boundary around $z\simeq2.1$ while preserving the already-strong \texttt{GALAXY}/\texttt{STAR} performance.

\begin{figure*}
\plotone{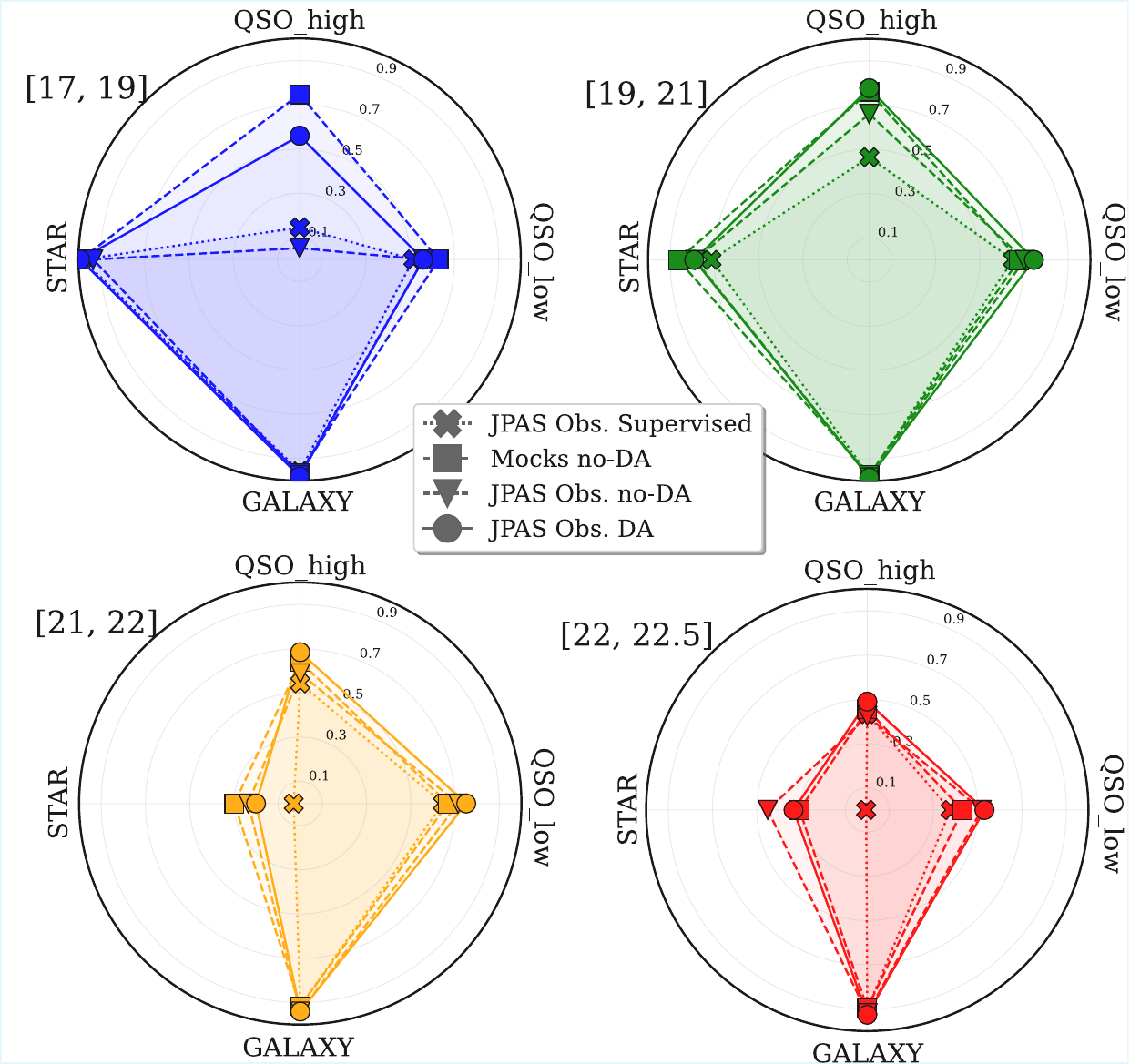}
\caption{
\textbf{Class-wise F1 vs.\ $r$-band magnitude cuts.}
Each polygon shows per--class F1 for a specific $r$-band selection (cuts indicated in the legend). 
Axes correspond to classes (\texttt{QSO\_high}, \texttt{QSO\_low}, \texttt{GALAXY}, \texttt{STAR}); the radius encodes F1 in $[0,1]$ with concentric gridlines for reference.
Markers and line-styles match Fig.~\ref{fig:F1_and_ROC_vert}.
For each magnitude cut, the J-PAS--based regimes are tested on the \emph{J-PAS Observations test split}, while the ``\emph{Mocks no-DA}'' pipeline is evaluated employing the \emph{mocks test split}.
Because the J-PAS$\times$DESI test set is comparatively small, some cuts leave few objects per class (e.g., only 11 \texttt{QSO\_high} for $17\!\le\!r\!<\!19$ and 54/28 \texttt{STAR} for $21\!\le\!r\!<\!22$ and $22\!<\!r\!\le\!22.5$), so apparent swings can be dominated by counting noise and should not be over-interpreted.
}
\label{fig:magnitudes_F1_radar}
\end{figure*}

Figure~\ref{fig:magnitudes_F1_radar} explores a simple operational knob for spectroscopic targeting: imposing $r$-band magnitude cuts and measuring how per-class F1 varies on the J-PAS$\times$DESI test split. Due to the size of the cross-match test set, several magnitude bins contain few objects per class---e.g., only 11 \texttt{QSO\_high} for $17\le r<19$, and 54/28 \texttt{STAR} for $21\le r<22$ and $22<r\le 22.5$---so apparent swings in F1 can be dominated by noise. With that caveat, the qualitative pattern is as expected: brighter cuts (higher S/N) tend to help both quasar subclasses. Across cuts, the SSDA model generally matches or improves upon the target-only baseline; in sparsely populated bins, uncertainties are large and differences should not be over-interpreted. Practically, we recommend selecting thresholds in the low--false-positive region of the ROC curves (Sect.~\ref{sec:results}) and aggregating magnitude bins---or enforcing a minimum per-class count---when using F1 to tune $r$ cuts. Because SSDA uses only a small number of J-PAS labels, the same procedure can be refreshed field-by-field as depth and seeing vary; storing selection probabilities then makes the targeting function explicit for downstream $n(z)$ and clustering analyses.


\bibliography{main}{}

\begin{thebibliography}{}
\expandafter\ifx\csname natexlab\endcsname\relax\def\natexlab#1{#1}\fi
\providecommand{\url}[1]{\href{#1}{#1}}
\providecommand{\dodoi}[1]{doi:~\href{http://doi.org/#1}{\nolinkurl{#1}}}
\providecommand{\doeprint}[1]{\href{http://ascl.net/#1}{\nolinkurl{http://ascl.net/#1}}}
\providecommand{\doarXiv}[1]{\href{https://arxiv.org/abs/#1}{\nolinkurl{https://arxiv.org/abs/#1}}}

\bibitem[{A.~G. {Adame} {et~al.}(2025){Adame}, {Aguilar}, {Ahlen}, {Alam}, {Alexander}, {Alvarez}, {Alves}, {Anand}, {Andrade}, {Armengaud}, {Avila}, {Aviles}, {Awan}, {Bahr-Kalus}, {Bailey}, {Baltay}, {Bault}, {Behera}, {BenZvi}, {Bera}, {Beutler}, {Bianchi}, {Blake}, {Blum}, {Brieden}, {Brodzeller}, {Brooks}, {Buckley-Geer}, {Burtin}, {Calderon}, {Canning}, {Carnero Rosell}, {Cereskaite}, {Cervantes-Cota}, {Chabanier}, {Chaussidon}, {Chaves-Montero}, {Chen}, {Chen}, {Claybaugh}, {Cole}, {Cuceu}, {Davis}, {Dawson}, {de la Macorra}, {de Mattia}, {Deiosso}, {Dey}, {Dey}, {Ding}, {Doel}, {Edelstein}, {Eftekharzadeh}, {Eisenstein}, {Elliott}, {Fagrelius}, {Fanning}, {Ferraro}, {Ereza}, {Findlay}, {Flaugher}, {Font-Ribera}, {Forero-S{\'a}nchez}, {Forero-Romero}, {Frenk}, {Garcia-Quintero}, {Gazta{\~n}aga}, {Gil-Mar{\'\i}n}, {Gontcho a Gontcho}, {Gonzalez-Morales}, {Gonzalez-Perez}, {Gordon}, {Green}, {Gruen}, {Gsponer}, {Gutierrez}, {Guy}, {Hadzhiyska}, {Hahn}, {Hanif}, {Herrera-Alcantar}, {Honscheid}, {Howlett},
  {Huterer}, {Ir{\v{s}}i{\v{c}}}, {Ishak}, {Juneau}, {Kara{\c{c}}ayl{\i}}, {Kehoe}, {Kent}, {Kirkby}, {Kremin}, {Krolewski}, {Lai}, {Lan}, {Landriau}, {Lang}, {Lasker}, {Le Goff}, {Le Guillou}, {Leauthaud}, {Levi}, {Li}, {Linder}, {Lodha}, {Magneville}, {Manera}, {Margala}, {Martini}, {Maus}, {McDonald}, {Medina-Varela}, {Meisner}, {Mena-Fern{\'a}ndez}, {Miquel}, {Moon}, {Moore}, {Moustakas}, {Mueller}, {Mu{\~n}oz-Guti{\'e}rrez}, {Myers}, {Nadathur}, {Napolitano}, {Neveux}, {Newman}, {Nguyen}, {Nie}, {Niz}, {Noriega}, {Padmanabhan}, {Paillas}, {Palanque-Delabrouille}, {Pan}, {Penmetsa}, {Percival}, {Pieri}, {Pinon}, {Poppett}, {Porredon}, {Prada}, {P{\'e}rez-Fern{\'a}ndez}, {P{\'e}rez-R{\`a}fols}, {Rabinowitz}, {Raichoor}, {Ram{\'\i}rez-P{\'e}rez}, {Ramirez-Solano}, {Rashkovetskyi}, {Ravoux}, {Rezaie}, {Rich}, {Rocher}, {Rockosi}, {Roe}, {Rosado-Marin}, {Ross}, {Rossi}, {Ruggeri}, {Ruhlmann-Kleider}, {Samushia}, {Sanchez}, {Saulder}, {Schlafly}, {Schlegel}, {Schubnell}, {Seo}, {Shafieloo}, {Sharples},
  {Silber}, {Slosar}, {Smith}, {Sprayberry}, {Tan}, {Tarl{\'e}}, {Taylor}, {Trusov}, {Ure{\~n}a-L{\'o}pez}, {Vaisakh}, {Valcin}, {Valdes}, {Vargas-Maga{\~n}a}, {Verde}, {Walther}, {Wang}, {Wang}, {Weaver}, {Weaverdyck}, {Wechsler}, {Weinberg}, {White}, {Yu}, {Yu}, {Yuan}, {Y{\`e}che}, {Zaborowski}, {Zarrouk}, {Zhang}, {Zhao}, {Zhao}, {Zhou}, \& {Zhuang}}]{2025JCAP...02..021A}
{Adame}, A.~G., {Aguilar}, J., {Ahlen}, S., {et~al.} 2025, \bibinfo{title}{{DESI 2024 VI: cosmological constraints from the measurements of baryon acoustic oscillations},} \jcap, 2025, 021, \dodoi{10.1088/1475-7516/2025/02/021}

\bibitem[{S. {Agarwal} {et~al.}(2024){Agarwal}, {{\'C}iprijanovi{\'c}}, \& {Nord}}]{2024arXiv241103334A}
{Agarwal}, S., {{\'C}iprijanovi{\'c}}, A., \& {Nord}, B.~D. 2024, \bibinfo{title}{{Neural Network Prediction of Strong Lensing Systems with Domain Adaptation and Uncertainty Quantification},} arXiv e-prints, arXiv:2411.03334, \dodoi{10.48550/arXiv.2411.03334}

\bibitem[{A. {Akhmetzhanova} {et~al.}(2025){Akhmetzhanova}, {Cuesta-Lazaro}, \& {Mishra-Sharma}}]{2025arXiv250805744A}
{Akhmetzhanova}, A., {Cuesta-Lazaro}, C., \& {Mishra-Sharma}, S. 2025, \bibinfo{title}{{Detecting Model Misspecification in Cosmology with Scale-Dependent Normalizing Flows},} arXiv e-prints, arXiv:2508.05744, \dodoi{10.48550/arXiv.2508.05744}

\bibitem[{A. {Akhmetzhanova} {et~al.}(2024){Akhmetzhanova}, {Mishra-Sharma}, \& {Dvorkin}}]{2024MNRAS.527.7459A}
{Akhmetzhanova}, A., {Mishra-Sharma}, S., \& {Dvorkin}, C. 2024, \bibinfo{title}{{Data compression and inference in cosmology with self-supervised machine learning},} \mnras, 527, 7459, \dodoi{10.1093/mnras/stad3646}

\bibitem[{S. {Alexander} {et~al.}(2023){Alexander}, {Gleyzer}, {Parul}, {Reddy}, {Tidball}, \& {Toomey}}]{2023ApJ...954...28A}
{Alexander}, S., {Gleyzer}, S., {Parul}, H., {et~al.} 2023, \bibinfo{title}{{Domain Adaptation for Simulation-based Dark Matter Searches with Strong Gravitational Lensing},} \apj, 954, 28, \dodoi{10.3847/1538-4357/acdfc7}

\bibitem[{S. {Alshammari} {et~al.}(2025){Alshammari}, {Hershey}, {Feldmann}, {Freeman}, \& {Hamilton}}]{2025arXiv250416929A}
{Alshammari}, S., {Hershey}, J., {Feldmann}, A., {Freeman}, W.~T., \& {Hamilton}, M. 2025, \bibinfo{title}{{I-Con: A Unifying Framework for Representation Learning},} arXiv e-prints, arXiv:2504.16929, \dodoi{10.48550/arXiv.2504.16929}

\bibitem[{N. {Anau Montel} {et~al.}(2023){Anau Montel}, {Coogan}, {Correa}, {Karchev}, \& {Weniger}}]{2023MNRAS.518.2746A}
{Anau Montel}, N., {Coogan}, A., {Correa}, C., {Karchev}, K., \& {Weniger}, C. 2023, \bibinfo{title}{{Estimating the warm dark matter mass from strong lensing images with truncated marginal neural ratio estimation},} \mnras, 518, 2746, \dodoi{10.1093/mnras/stac3215}

\bibitem[{S. {Andrianomena} \& S. {Hassan}(2025){Andrianomena} \& {Hassan}}]{2025Ap&SS.370...14A}
{Andrianomena}, S., \& {Hassan}, S. 2025, \bibinfo{title}{{Towards cosmological inference on unlabeled out-of-distribution HI observational data},} \apss, 370, 14, \dodoi{10.1007/s10509-025-04405-y}

\bibitem[{P.~O. {Baqui} {et~al.}(2021){Baqui}, {Marra}, {Casarini}, {Angulo}, {D{\'\i}az-Garc{\'\i}a}, {Hern{\'a}ndez-Monteagudo}, {Lopes}, {L{\'o}pez-Sanjuan}, {Muniesa}, {Placco}, {Quartin}, {Queiroz}, {Sobral}, {Solano}, {Tempel}, {Varela}, {V{\'\i}lchez}, {Abramo}, {Alcaniz}, {Benitez}, {Bonoli}, {Carneiro}, {Cenarro}, {Crist{\'o}bal-Hornillos}, {de Amorim}, {de Oliveira}, {Dupke}, {Ederoclite}, {Gonz{\'a}lez Delgado}, {Mar{\'\i}n-Franch}, {Moles}, {V{\'a}zquez Rami{\'o}}, {Sodr{\'e}}, \& {Taylor}}]{2021A&A...645A..87B}
{Baqui}, P.~O., {Marra}, V., {Casarini}, L., {et~al.} 2021, \bibinfo{title}{{The miniJPAS survey: star-galaxy classification using machine learning},} \aap, 645, A87, \dodoi{10.1051/0004-6361/202038986}

\bibitem[{A. {Bardes} {et~al.}(2021){Bardes}, {Ponce}, \& {LeCun}}]{2021arXiv210504906B}
{Bardes}, A., {Ponce}, J., \& {LeCun}, Y. 2021, \bibinfo{title}{{VICReg: Variance-Invariance-Covariance Regularization for Self-Supervised Learning},} arXiv e-prints, arXiv:2105.04906, \dodoi{10.48550/arXiv.2105.04906}

\bibitem[{N. {Benitez} {et~al.}(2014){Benitez}, {Dupke}, {Moles}, {Sodre}, {Cenarro}, {Marin-Franch}, {Taylor}, {Cristobal}, {Fernandez-Soto}, {Mendes de Oliveira}, {Cepa-Nogue}, {Abramo}, {Alcaniz}, {Overzier}, {Hernandez-Monteagudo}, {Alfaro}, {Kanaan}, {Carvano}, {Reis}, {Martinez Gonzalez}, {Ascaso}, {Ballesteros}, {Xavier}, {Varela}, {Ederoclite}, {Vazquez Ramio}, {Broadhurst}, {Cypriano}, {Angulo}, {Diego}, {Zandivarez}, {Diaz}, {Melchior}, {Umetsu}, {Spinelli}, {Zitrin}, {Coe}, {Yepes}, {Vielva}, {Sahni}, {Marcos-Caballero}, {Kitaura}, {Maroto}, {Masip}, {Tsujikawa}, {Carneiro}, {Gonzalez Nuevo}, {Carvalho}, {Reboucas}, {Carvalho}, {Abdalla}, {Bernui}, {Pigozzo}, {Ferreira}, {Chandrachani Devi}, {Bengaly}, {Campista}, {Amorim}, {Asari}, {Bongiovanni}, {Bonoli}, {Bruzual}, {Cardiel}, {Cava}, {Cid Fernandes}, {Coelho}, {Cortesi}, {Delgado}, {Diaz Garcia}, {Espinosa}, {Galliano}, {Gonzalez-Serrano}, {Falcon-Barroso}, {Fritz}, {Fernandes}, {Gorgas}, {Hoyos}, {Jimenez-Teja}, {Lopez-Aguerri}, {Lopez-San
  Juan}, {Mateus}, {Molino}, {Novais}, {OMill}, {Oteo}, {Perez-Gonzalez}, {Poggianti}, {Proctor}, {Ricciardelli}, {Sanchez-Blazquez}, {Storchi-Bergmann}, {Telles}, {Schoennell}, {Trujillo}, {Vazdekis}, {Viironen}, {Daflon}, {Aparicio-Villegas}, {Rocha}, {Ribeiro}, {Borges}, {Martins}, {Marcolino}, {Martinez-Delgado}, {Perez-Torres}, {Siffert}, {Calvao}, {Sako}, {Kessler}, {Alvarez-Candal}, {De Pra}, {Roig}, {Lazzaro}, {Gorosabel}, {Lopes de Oliveira}, {Lima-Neto}, {Irwin}, {Liu}, {Alvarez}, {Balmes}, {Chueca}, {Costa-Duarte}, {da Costa}, {Dantas}, {Diaz}, {Fabregat}, {Ferrari}, {Gavela}, {Gracia}, {Gruel}, {Gutierrez}, {Guzman}, {Hernandez-Fernandez}, {Herranz}, {Hurtado-Gil}, {Jablonsky}, {Laporte}, {Le Tiran}, {Licandro}, {Lima}, {Martin}, {Martinez}, {Montero}, {Penteado}, {Pereira}, {Peris}, {Quilis}, {Sanchez-Portal}, {Soja}, {Solano}, {Torra}, \& {Valdivielso}}]{2014arXiv1403.5237B}
{Benitez}, N., {Dupke}, R., {Moles}, M., {et~al.} 2014, \bibinfo{title}{{J-PAS: The Javalambre-Physics of the Accelerated Universe Astrophysical Survey},} arXiv e-prints, arXiv:1403.5237, \dodoi{10.48550/arXiv.1403.5237}

\bibitem[{P. {Bergamini} {et~al.}(2023){Bergamini}, {Grillo}, {Rosati}, {Vanzella}, {Me{\v{s}}tri{\'c}}, {Mercurio}, {Acebron}, {Caminha}, {Granata}, {Meneghetti}, {Angora}, \& {Nonino}}]{2023A&A...674A..79B}
{Bergamini}, P., {Grillo}, C., {Rosati}, P., {et~al.} 2023, \bibinfo{title}{{A state-of-the-art strong-lensing model of MACS J0416.1{\ensuremath{-}}2403 with the largest sample of spectroscopic multiple images},} \aap, 674, A79, \dodoi{10.1051/0004-6361/202244834}

\bibitem[{D. {Berthelot} {et~al.}(2021){Berthelot}, {Roelofs}, {Sohn}, {Carlini}, \& {Kurakin}}]{2021arXiv210604732B}
{Berthelot}, D., {Roelofs}, R., {Sohn}, K., {Carlini}, N., \& {Kurakin}, A. 2021, \bibinfo{title}{{AdaMatch: A Unified Approach to Semi-Supervised Learning and Domain Adaptation},} arXiv e-prints, arXiv:2106.04732, \dodoi{10.48550/arXiv.2106.04732}

\bibitem[{E. {Bertin} \& S. {Arnouts}(1996){Bertin} \& {Arnouts}}]{1996A&AS..117..393B}
{Bertin}, E., \& {Arnouts}, S. 1996, \bibinfo{title}{{SExtractor: Software for source extraction.},} \aaps, 117, 393, \dodoi{10.1051/aas:1996164}

\bibitem[{L. Biewald(2020)Biewald}]{wandb}
Biewald, L. 2020, Experiment Tracking with Weights and Biases, \url{https://www.wandb.com/}

\bibitem[{T. {Chen} {et~al.}(2020){Chen}, {Kornblith}, {Norouzi}, \& {Hinton}}]{2020arXiv200205709C}
{Chen}, T., {Kornblith}, S., {Norouzi}, M., \& {Hinton}, G. 2020, \bibinfo{title}{{A Simple Framework for Contrastive Learning of Visual Representations},} arXiv e-prints, arXiv:2002.05709, \dodoi{10.48550/arXiv.2002.05709}

\bibitem[{A. {{\'C}iprijanovi{\'c}} {et~al.}(2023){{\'C}iprijanovi{\'c}}, {Lewis}, {Pedro}, {Madireddy}, {Nord}, {Perdue}, \& {Wild}}]{2023MLS&T...4b5013C}
{{\'C}iprijanovi{\'c}}, A., {Lewis}, A., {Pedro}, K., {et~al.} 2023, \bibinfo{title}{{DeepAstroUDA: semi-supervised universal domain adaptation for cross-survey galaxy morphology classification and anomaly detection},} Machine Learning: Science and Technology, 4, 025013, \dodoi{10.1088/2632-2153/acca5f}

\bibitem[{A. {{\'C}iprijanovi{\'c}} {et~al.}(2021){{\'C}iprijanovi{\'c}}, {Kafkes}, {Downey}, {Jenkins}, {Perdue}, {Madireddy}, {Johnston}, {Snyder}, \& {Nord}}]{2021MNRAS.506..677C}
{{\'C}iprijanovi{\'c}}, A., {Kafkes}, D., {Downey}, K., {et~al.} 2021, \bibinfo{title}{{DeepMerge - II. Building robust deep learning algorithms for merging galaxy identification across domains},} \mnras, 506, 677, \dodoi{10.1093/mnras/stab1677}

\bibitem[{R.~A. {Crain} {et~al.}(2015){Crain}, {Schaye}, {Bower}, {Furlong}, {Schaller}, {Theuns}, {Dalla Vecchia}, {Frenk}, {McCarthy}, {Helly}, {Jenkins}, {Rosas-Guevara}, {White}, \& {Trayford}}]{2015MNRAS.450.1937C}
{Crain}, R.~A., {Schaye}, J., {Bower}, R.~G., {et~al.} 2015, \bibinfo{title}{{The EAGLE simulations of galaxy formation: calibration of subgrid physics and model variations},} \mnras, 450, 1937, \dodoi{10.1093/mnras/stv725}

\bibitem[{G. {Dalton} {et~al.}(2016){Dalton}, {Trager}, {Abrams}, {Bonifacio}, {Aguerri}, {Middleton}, {Benn}, {Dee}, {Say{\`e}de}, {Lewis}, {Pragt}, {Pico}, {Walton}, {Rey}, {Allende Prieto}, {Pe{\~n}ate}, {Lhome}, {Ag{\'o}cs}, {Alonso}, {Terrett}, {Brock}, {Gilbert}, {Schallig}, {Ridings}, {Guinouard}, {Verheijen}, {Tosh}, {Rogers}, {Lee}, {Steele}, {Stuik}, {Tromp}, {Jask{\'o}}, {Carrasco}, {Farcas}, {Kragt}, {Lesman}, {Kroes}, {Mottram}, {Bates}, {Rodriguez}, {Gribbin}, {Delgado}, {Herreros}, {Martin}, {Cano}, {Navarro}, {Irwin}, {Lewis}, {Gonzalez Solares}, {Murphy}, {Worley}, {Bassom}, {O'Mahoney}, {Bianco}, {Zurita}, {ter Horst}, {Molinari}, {Lodi}, {Guerra}, {Martin}, {Vallenari}, {Salasnich}, {Baruffolo}, {Jin}, {Hill}, {Smith}, {Drew}, {Poggianti}, {Pieri}, {Dominquez Palmero}, \& {Farina}}]{2016SPIE.9908E..1GD}
{Dalton}, G., {Trager}, S., {Abrams}, D.~C., {et~al.} 2016, \bibinfo{title}{{Final design and progress of WEAVE: the next generation wide-field spectroscopy facility for the William Herschel Telescope},} in Society of Photo-Optical Instrumentation Engineers (SPIE) Conference Series, Vol. 9908, Ground-based and Airborne Instrumentation for Astronomy VI, ed. C.~J. {Evans}, L.~{Simard}, \& H.~{Takami}, 99081G, \dodoi{10.1117/12.2231078}

\bibitem[{A. {del Pino} {et~al.}(2024){del Pino}, {L{\'o}pez-Sanjuan}, {Hern{\'a}n-Caballero}, {Dom{\'\i}nguez-S{\'a}nchez}, {von Marttens}, {Fern{\'a}ndez-Ontiveros}, {Coelho}, {Lumbreras-Calle}, {Vega-Ferrero}, {Jimenez-Esteban}, {Cruz}, {Marra}, {Quartin}, {Galarza}, {Angulo}, {Cenarro}, {Crist{\'o}bal-Hornillos}, {Dupke}, {Ederoclite}, {Hern{\'a}ndez-Monteagudo}, {Mar{\'\i}n-Franch}, {Moles}, {Sodr{\'e}}, {Varela}, \& {V{\'a}zquez Rami{\'o}}}]{2024A&A...691A.221D}
{del Pino}, A., {L{\'o}pez-Sanjuan}, C., {Hern{\'a}n-Caballero}, A., {et~al.} 2024, \bibinfo{title}{{J-PLUS: Bayesian object classification with a strum of BANNJOS},} \aap, 691, A221, \dodoi{10.1051/0004-6361/202450503}

\bibitem[{ {DESI Collaboration} {et~al.}(2024){DESI Collaboration}, {Adame}, {Aguilar}, {Ahlen}, {Alam}, {Aldering}, {Alexander}, {Alfarsy}, {Allende Prieto}, {Alvarez}, {Alves}, {Anand}, {Andrade-Oliveira}, {Armengaud}, {Asorey}, {Avila}, {Aviles}, {Bailey}, {Balaguera-Antol{\'\i}nez}, {Ballester}, {Baltay}, {Bault}, {Bautista}, {Behera}, {Beltran}, {BenZvi}, {Beraldo e Silva}, {Bermejo-Climent}, {Berti}, {Besuner}, {Beutler}, {Bianchi}, {Blake}, {Blum}, {Bolton}, {Brieden}, {Brodzeller}, {Brooks}, {Brown}, {Buckley-Geer}, {Burtin}, {Cabayol-Garcia}, {Cai}, {Canning}, {Cardiel-Sas}, {Carnero Rosell}, {Castander}, {Cervantes-Cota}, {Chabanier}, {Chaussidon}, {Chaves-Montero}, {Chen}, {Chen}, {Chuang}, {Claybaugh}, {Cole}, {Cooper}, {Cuceu}, {Davis}, {Dawson}, {de Belsunce}, {de la Cruz}, {de la Macorra}, {Della Costa}, {de Mattia}, {Demina}, {Demirbozan}, {DeRose}, {Dey}, {Dey}, {Dhungana}, {Ding}, {Ding}, {Doel}, {Doshi}, {Douglass}, {Edge}, {Eftekharzadeh}, {Eisenstein}, {Elliott}, {Ereza}, {Escoffier},
  {Fagrelius}, {Fan}, {Fanning}, {Fawcett}, {Ferraro}, {Flaugher}, {Font-Ribera}, {Forero-Romero}, {Forero-S{\'a}nchez}, {Frenk}, {G{\"a}nsicke}, {Garc{\'\i}a}, {Garc{\'\i}a-Bellido}, {Garcia-Quintero}, {Garrison}, {Gil-Mar{\'\i}n}, {Golden-Marx}, {Gontcho A Gontcho}, {Gonzalez-Morales}, {Gonzalez-Perez}, {Gordon}, {Graur}, {Green}, {Gruen}, {Guy}, {Hadzhiyska}, {Hahn}, {Han}, {Hanif}, {Herrera-Alcantar}, {Honscheid}, {Hou}, {Howlett}, {Huterer}, {Ir{\v{s}}i{\v{c}}}, {Ishak}, {Jacques}, {Jana}, {Jiang}, {Jimenez}, {Jing}, {Joudaki}, {Joyce}, {Jullo}, {Juneau}, {Kara{\c{c}}ayl{\i}}, {Karim}, {Kehoe}, {Kent}, {Khederlarian}, {Kim}, {Kirkby}, {Kisner}, {Kitaura}, {Kizhuprakkat}, {Kneib}, {Koposov}, {Kov{\'a}cs}, {Kremin}, {Krolewski}, {L'Huillier}, {Lahav}, {Lambert}, {Lamman}, {Lan}, {Landriau}, {Lang}, {Lange}, {Lasker}, {Leauthaud}, {Le Guillou}, {Levi}, {Li}, {Linder}, {Lyons}, {Magneville}, {Manera}, {Manser}, {Margala}, {Martini}, {McDonald}, {Medina}, {Medina-Varela}, {Meisner}, {Mena-Fern{\'a}ndez},
  {Meneses-Rizo}, {Mezcua}, {Miquel}, {Montero-Camacho}, {Moon}, {Moore}, {Moustakas}, {Mueller}, {Mundet}, {Mu{\~n}oz-Guti{\'e}rrez}, {Myers}, {Nadathur}, {Napolitano}, {Neveux}, {Newman}, {Nie}, {Nikutta}, {Niz}, {Norberg}, {Noriega}, {Paillas}, {Palanque-Delabrouille}, {Palmese}, {Pan}, {Parkinson}, {Penmetsa}, {Percival}, {P{\'e}rez-Fern{\'a}ndez}, {P{\'e}rez-R{\`a}fols}, {Pieri}, {Poppett}, {Porredon}, \& {Pothier}}]{2024AJ....168...58D}
{DESI Collaboration}, {Adame}, A.~G., {Aguilar}, J., {et~al.} 2024, \bibinfo{title}{{The Early Data Release of the Dark Energy Spectroscopic Instrument},} \aj, 168, 58, \dodoi{10.3847/1538-3881/ad3217}

\bibitem[{ {Euclid Collaboration} {et~al.}(2025){Euclid Collaboration}, {Mellier}, {Abdurro'uf}, {Acevedo Barroso}, {Ach{\'u}carro}, {Adamek}, {Adam}, {Addison}, {Aghanim}, {Aguena}, {Ajani}, {Akrami}, {Al-Bahlawan}, {Alavi}, {Albuquerque}, {Alestas}, {Alguero}, {Allaoui}, {Allen}, {Allevato}, {Alonso-Tetilla}, {Altieri}, {Alvarez-Candal}, {Alvi}, {Amara}, {Amendola}, {Amiaux}, {Andika}, {Andreon}, {Andrews}, {Angora}, {Angulo}, {Annibali}, {Anselmi}, {Anselmi}, {Arcari}, {Archidiacono}, {Aric{\`o}}, {Arnaud}, {Arnouts}, {Asgari}, {Asorey}, {Atayde}, {Atek}, {Atrio-Barandela}, {Aubert}, {Aubourg}, {Auphan}, {Auricchio}, {Aussel}, {Aussel}, {Avelino}, {Avgoustidis}, {Avila}, {Awan}, {Azzollini}, {Baccigalupi}, {Bachelet}, {Bacon}, {Baes}, {Bagley}, {Bahr-Kalus}, {Balaguera-Antolinez}, {Balbinot}, {Balcells}, {Baldi}, {Baldry}, {Balestra}, {Ballardini}, {Ballester}, {Balogh}, {Ba{\~n}ados}, {Barbier}, {Bardelli}, {Baron}, {Barreiro}, {Barrena}, {Barriere}, {Barros}, {Barthelemy}, {Bartolo}, {Basset},
  {Battaglia}, {Battisti}, {Baugh}, {Baumont}, {Bazzanini}, {Beaulieu}, {Beckmann}, {Belikov}, {Bel}, {Bellagamba}, {Bella}, {Bellini}, {Benabed}, {Bender}, {Benevento}, {Bennett}, {Benson}, {Bergamini}, {Bermejo-Climent}, {Bernardeau}, {Bertacca}, {Berthe}, {Berthier}, {Bethermin}, {Beutler}, {Bevillon}, {Bhargava}, {Bhatawdekar}, {Bianchi}, {Bisigello}, {Biviano}, {Blake}, {Blanchard}, {Blazek}, {Blot}, {Bosco}, {Bodendorf}, {Boenke}, {B{\"o}hringer}, {Boldrini}, {Bolzonella}, {Bonchi}, {Bonici}, {Bonino}, {Bonino}, {Bonvin}, {Bon}, {Booth}, {Borgani}, {Borlaff}, {Borsato}, {Bose}, {Botticella}, {Boucaud}, {Bouche}, {Boucher}, {Boutigny}, {Bouvard}, {Bouwens}, {Bouy}, {Bowler}, {Bozza}, {Bozzo}, {Branchini}, {Brando}, {Brau-Nogue}, {Brekke}, {Bremer}, {Brescia}, {Breton}, {Brinchmann}, {Brinckmann}, {Brockley-Blatt}, {Brodwin}, {Brouard}, {Brown}, {Bruton}, {Bucko}, {Buddelmeijer}, {Buenadicha}, {Buitrago}, {Burger}, {Burigana}, {Busillo}, {Busonero}, {Cabanac}, {Cabayol-Garcia}, {Cagliari}, {Caillat},
  {Caillat}, {Calabrese}, {Calabro}, {Calderone}, {Calura}, {Camacho Quevedo}, {Camera}, {Campos}, {Ca{\~n}as-Herrera}, {Candini}, {Cantiello}, {Capobianco}, {Cappellaro}, {Cappelluti}, {Cappi}, {Caputi}, {Cara}, {Carbone}, {Cardone}, {Carella}, {Carlberg}, {Carle}, {Carminati}, {Caro}, {Carrasco}, {Carretero}, {Carrilho}, {Carron Duque}, \& {Carry}}]{2025A&A...697A...1E}
{Euclid Collaboration}, {Mellier}, Y., {Abdurro'uf}, {et~al.} 2025, \bibinfo{title}{{Euclid: I. Overview of the Euclid mission},} \aap, 697, A1, \dodoi{10.1051/0004-6361/202450810}

\bibitem[{Y. {Feng} {et~al.}(2016){Feng}, {Chu}, {Seljak}, \& {McDonald}}]{2016MNRAS.463.2273F}
{Feng}, Y., {Chu}, M.-Y., {Seljak}, U., \& {McDonald}, P. 2016, \bibinfo{title}{{FASTPM: a new scheme for fast simulations of dark matter and haloes},} \mnras, 463, 2273, \dodoi{10.1093/mnras/stw2123}

\bibitem[{Y. {Ganin} {et~al.}(2015){Ganin}, {Ustinova}, {Ajakan}, {Germain}, {Larochelle}, {Laviolette}, {Marchand}, \& {Lempitsky}}]{2015arXiv150507818G}
{Ganin}, Y., {Ustinova}, E., {Ajakan}, H., {et~al.} 2015, \bibinfo{title}{{Domain-Adversarial Training of Neural Networks},} arXiv e-prints, arXiv:1505.07818, \dodoi{10.48550/arXiv.1505.07818}

\bibitem[{L.~H. {Garrison} {et~al.}(2021){Garrison}, {Eisenstein}, {Ferrer}, {Maksimova}, \& {Pinto}}]{2021MNRAS.508..575G}
{Garrison}, L.~H., {Eisenstein}, D.~J., {Ferrer}, D., {Maksimova}, N.~A., \& {Pinto}, P.~A. 2021, \bibinfo{title}{{The ABACUS cosmological N-body code},} \mnras, 508, 575, \dodoi{10.1093/mnras/stab2482}

\bibitem[{S. {Gilda} {et~al.}(2024){Gilda}, {de Mathelin}, {Bellstedt}, \& {Richard}}]{2024Astro...3..189G}
{Gilda}, S., {de Mathelin}, A., {Bellstedt}, S., \& {Richard}, G. 2024, \bibinfo{title}{{Unsupervised Domain Adaptation for Constraining Star Formation Histories},} Astronomy, 3, 189, \dodoi{10.3390/astronomy3030012}

\bibitem[{Z. {Han} {et~al.}(2024){Han}, {Gao}, {Liu}, {Zhang}, \& {Qian Zhang}}]{2024arXiv240314608H}
{Han}, Z., {Gao}, C., {Liu}, J., {Zhang}, J., \& {Qian Zhang}, S. 2024, \bibinfo{title}{{Parameter-Efficient Fine-Tuning for Large Models: A Comprehensive Survey},} arXiv e-prints, arXiv:2403.14608, \dodoi{10.48550/arXiv.2403.14608}

\bibitem[{M.~A. {Hayat} {et~al.}(2021){Hayat}, {Stein}, {Harrington}, {Luki{\'c}}, \& {Mustafa}}]{2021ApJ...911L..33H}
{Hayat}, M.~A., {Stein}, G., {Harrington}, P., {Luki{\'c}}, Z., \& {Mustafa}, M. 2021, \bibinfo{title}{{Self-supervised Representation Learning for Astronomical Images},} \apjl, 911, L33, \dodoi{10.3847/2041-8213/abf2c7}

\bibitem[{M. {Huertas-Company} {et~al.}(2023){Huertas-Company}, {Sarmiento}, \& {Knapen}}]{2023RASTI...2..441H}
{Huertas-Company}, M., {Sarmiento}, R., \& {Knapen}, J.~H. 2023, \bibinfo{title}{{A brief review of contrastive learning applied to astrophysics},} RAS Techniques and Instruments, 2, 441, \dodoi{10.1093/rasti/rzad028}

\bibitem[{J.~D. Hunter(2007)Hunter}]{Matplotlib-Hunter07}
Hunter, J.~D. 2007, \bibinfo{title}{Matplotlib: A 2D graphics environment,} Computing in Science \& Engineering, 9, 90, \dodoi{10.1109/MCSE.2007.55}

\bibitem[{{\v{Z}}. {Ivezi{\'c}} {et~al.}(2019){Ivezi{\'c}}, {Kahn}, {Tyson}, {Abel}, {Acosta}, {Allsman}, {Alonso}, {AlSayyad}, {Anderson}, {Andrew}, {Angel}, {Angeli}, {Ansari}, {Antilogus}, {Araujo}, {Armstrong}, {Arndt}, {Astier}, {Aubourg}, {Auza}, {Axelrod}, {Bard}, {Barr}, {Barrau}, {Bartlett}, {Bauer}, {Bauman}, {Baumont}, {Bechtol}, {Bechtol}, {Becker}, {Becla}, {Beldica}, {Bellavia}, {Bianco}, {Biswas}, {Blanc}, {Blazek}, {Blandford}, {Bloom}, {Bogart}, {Bond}, {Booth}, {Borgland}, {Borne}, {Bosch}, {Boutigny}, {Brackett}, {Bradshaw}, {Brandt}, {Brown}, {Bullock}, {Burchat}, {Burke}, {Cagnoli}, {Calabrese}, {Callahan}, {Callen}, {Carlin}, {Carlson}, {Chandrasekharan}, {Charles-Emerson}, {Chesley}, {Cheu}, {Chiang}, {Chiang}, {Chirino}, {Chow}, {Ciardi}, {Claver}, {Cohen-Tanugi}, {Cockrum}, {Coles}, {Connolly}, {Cook}, {Cooray}, {Covey}, {Cribbs}, {Cui}, {Cutri}, {Daly}, {Daniel}, {Daruich}, {Daubard}, {Daues}, {Dawson}, {Delgado}, {Dellapenna}, {de Peyster}, {de Val-Borro}, {Digel}, {Doherty},
  {Dubois}, {Dubois-Felsmann}, {Durech}, {Economou}, {Eifler}, {Eracleous}, {Emmons}, {Fausti Neto}, {Ferguson}, {Figueroa}, {Fisher-Levine}, {Focke}, {Foss}, {Frank}, {Freemon}, {Gangler}, {Gawiser}, {Geary}, {Gee}, {Geha}, {Gessner}, {Gibson}, {Gilmore}, {Glanzman}, {Glick}, {Goldina}, {Goldstein}, {Goodenow}, {Graham}, {Gressler}, {Gris}, {Guy}, {Guyonnet}, {Haller}, {Harris}, {Hascall}, {Haupt}, {Hernandez}, {Herrmann}, {Hileman}, {Hoblitt}, {Hodgson}, {Hogan}, {Howard}, {Huang}, {Huffer}, {Ingraham}, {Innes}, {Jacoby}, {Jain}, {Jammes}, {Jee}, {Jenness}, {Jernigan}, {Jevremovi{\'c}}, {Johns}, {Johnson}, {Johnson}, {Jones}, {Juramy-Gilles}, {Juri{\'c}}, {Kalirai}, {Kallivayalil}, {Kalmbach}, {Kantor}, {Karst}, {Kasliwal}, {Kelly}, {Kessler}, {Kinnison}, {Kirkby}, {Knox}, {Kotov}, {Krabbendam}, {Krughoff}, {Kub{\'a}nek}, {Kuczewski}, {Kulkarni}, {Ku}, {Kurita}, {Lage}, {Lambert}, {Lange}, {Langton}, {Le Guillou}, {Levine}, {Liang}, {Lim}, {Lintott}, {Long}, {Lopez}, {Lotz}, {Lupton}, {Lust}, {MacArthur},
  {Mahabal}, {Mandelbaum}, {Markiewicz}, {Marsh}, {Marshall}, {Marshall}, {May}, {McKercher}, {McQueen}, {Meyers}, {Migliore}, {Miller}, \& {Mills}}]{2019ApJ...873..111I}
{Ivezi{\'c}}, {\v{Z}}., {Kahn}, S.~M., {Tyson}, J.~A., {et~al.} 2019, \bibinfo{title}{{LSST: From Science Drivers to Reference Design and Anticipated Data Products},} \apj, 873, 111, \dodoi{10.3847/1538-4357/ab042c}

\bibitem[{A.~P. {Jeakel} {et~al.}(2025){Jeakel}, {Vieira dos Santos}, {Marra}, {von Marttens}, {Gurung-L{\'o}pez}, {Abramo}, {Alcaniz}, {Benitez}, {Bonoli}, {Cenarro}, {Crist{\'o}bal-Hornillos}, {Daflon}, {Dupke}, {Ederoclite}, {Gonz{\'a}lez Delgado}, {Hern{\'a}n-Caballero}, {Hern{\'a}ndez-Monteagudo}, {Liu}, {L{\'o}pez-Sanjuan}, {Mar{\'\i}n-Franch}, {Mendes de Oliveira}, {Moles}, {Roig}, {Sodr{\'e}}, {Taylor}, {Varela}, {V{\'a}zquez Rami{\'o}}, {Vilchez}, {Willmer}, \& {Zaragoza-Cardiel}}]{2025arXiv251120524J}
{Jeakel}, A.~P., {Vieira dos Santos}, G., {Marra}, V., {et~al.} 2025, \bibinfo{title}{{The miniJPAS and J-NEP surveys: Machine learning for star-galaxy separation},} arXiv e-prints, arXiv:2511.20524, \dodoi{10.48550/arXiv.2511.20524}

\bibitem[{S. {Jin} {et~al.}(2024){Jin}, {Trager}, {Dalton}, {Aguerri}, {Drew}, {Falc{\'o}n-Barroso}, {G{\"a}nsicke}, {Hill}, {Iovino}, {Pieri}, {Poggianti}, {Smith}, {Vallenari}, {Abrams}, {Aguado}, {Antoja}, {Arag{\'o}n-Salamanca}, {Ascasibar}, {Babusiaux}, {Balcells}, {Barrena}, {Battaglia}, {Belokurov}, {Bensby}, {Bonifacio}, {Bragaglia}, {Carrasco}, {Carrera}, {Cornwell}, {Dom{\'\i}nguez-Palmero}, {Duncan}, {Famaey}, {Fari{\~n}a}, {Gonzalez}, {Guest}, {Hatch}, {Hess}, {Hoskin}, {Irwin}, {Knapen}, {Koposov}, {Kuchner}, {Laigle}, {Lewis}, {Longhetti}, {Lucatello}, {M{\'e}ndez-Abreu}, {Mercurio}, {Molaeinezhad}, {Mongui{\'o}}, {Morrison}, {Murphy}, {Peralta de Arriba}, {P{\'e}rez}, {P{\'e}rez-R{\`a}fols}, {Pic{\'o}}, {Raddi}, {Romero-G{\'o}mez}, {Royer}, {Siebert}, {Seabroke}, {Som}, {Terrett}, {Thomas}, {Wesson}, {Worley}, {Alfaro}, {Allende Prieto}, {Alonso-Santiago}, {Amos}, {Ashley}, {Balaguer-N{\'u}{\~n}ez}, {Balbinot}, {Bellazzini}, {Benn}, {Berlanas}, {Bernard}, {Best}, {Bettoni}, {Bianco}, {Bishop},
  {Blomqvist}, {Boeche}, {Bolzonella}, {Bonoli}, {Bosma}, {Britavskiy}, {Busarello}, {Caffau}, {Cantat-Gaudin}, {Castro-Ginard}, {Couto}, {Carbajo-Hijarrubia}, {Carter}, {Casamiquela}, {Conrado}, {Corcho-Caballero}, {Costantin}, {Deason}, {de Burgos}, {De Grandi}, {Di Matteo}, {Dom{\'\i}nguez-G{\'o}mez}, {Dorda}, {Drake}, {Dutta}, {Erkal}, {Feltzing}, {Ferr{\'e}-Mateu}, {Feuillet}, {Figueras}, {Fossati}, {Franciosini}, {Frasca}, {Fumagalli}, {Gallazzi}, {Garc{\'\i}a-Benito}, {Gentile Fusillo}, {Gebran}, {Gilbert}, {Gledhill}, {Gonz{\'a}lez Delgado}, {Greimel}, {Guarcello}, {Guerra}, {Gullieuszik}, {Haines}, {Hardcastle}, {Harris}, {Haywood}, {Helmi}, {Hernandez}, {Herrero}, {Hughes}, {Ir{\v{s}}i{\v{c}}}, {Jablonka}, {Jarvis}, {Jordi}, {Kondapally}, {Kordopatis}, {Krogager}, {La Barbera}, {Lam}, {Larsen}, {Lemasle}, {Lewis}, {Lhom{\'e}}, {Lind}, {Lodi}, {Longobardi}, {Lonoce}, {Magrini}, {Ma{\'\i}z Apell{\'a}niz}, {Marchal}, {Marco}, {Martin}, {Matsuno}, {Maurogordato}, {Merluzzi}, {Miralda-Escud{\'e}},
  {Molinari}, {Monari}, {Morelli}, {Mottram}, {Naylor}, {Negueruela}, {O{\~n}orbe}, {Pancino}, {Peirani}, {Peletier}, {Pozzetti}, {Rainer}, {Ramos}, {Read}, {Rossi}, {R{\"o}ttgering}, {Rubi{\~n}o-Mart{\'\i}n}, {Sabater}, {San Juan}, {Sanna}, {Schallig}, {Schiavon}, {Schultheis}, {Serra}, {Shimwell}, {Sim{\'o}n-D{\'\i}az}, {Smith}, {Sordo}, {Sorini}, {Soubiran}, {Starkenburg}, {Steele}, {Stott}, {Stuik}, {Tolstoy}, {Tortora}, {Tsantaki}, {Van der Swaelmen}, {van Weeren}, \& {Vergani}}]{2024MNRAS.530.2688J}
{Jin}, S., {Trager}, S.~C., {Dalton}, G.~B., {et~al.} 2024, \bibinfo{title}{{The wide-field, multiplexed, spectroscopic facility WEAVE: Survey design, overview, and simulated implementation},} \mnras, 530, 2688, \dodoi{10.1093/mnras/stad557}

\bibitem[{D. {Lang} {et~al.}(2016){Lang}, {Hogg}, \& {Mykytyn}}]{2016ascl.soft04008L}
{Lang}, D., {Hogg}, D.~W., \& {Mykytyn}, D. 2016, {The Tractor: Probabilistic astronomical source detection and measurement},, Astrophysics Source Code Library, record ascl:1604.008 \doeprint{1604.008}

\bibitem[{J.-Y. {Lee} {et~al.}(2024){Lee}, {Kim}, {Jung}, {Oh}, {Jo}, {Park}, {Lee}, {Ting}, \& {Hwang}}]{2024ApJ...975...38L}
{Lee}, J.-Y., {Kim}, J.-h., {Jung}, M., {et~al.} 2024, \bibinfo{title}{{Inferring Cosmological Parameters on SDSS via Domain-generalized Neural Networks and Light-cone Simulations},} \apj, 975, 38, \dodoi{10.3847/1538-4357/ad73d4}

\bibitem[{C.-L. {Li} {et~al.}(2017){Li}, {Chang}, {Cheng}, {Yang}, \& {P{\'o}czos}}]{2017arXiv170508584L}
{Li}, C.-L., {Chang}, W.-C., {Cheng}, Y., {Yang}, Y., \& {P{\'o}czos}, B. 2017, \bibinfo{title}{{MMD GAN: Towards Deeper Understanding of Moment Matching Network},} arXiv e-prints, arXiv:1705.08584, \dodoi{10.48550/arXiv.1705.08584}

\bibitem[{X. {Liu} {et~al.}(2022){Liu}, {Yoo}, {Xing}, {Oh}, {El Fakhri}, {Kang}, \& {Woo}}]{2022arXiv220807422L}
{Liu}, X., {Yoo}, C., {Xing}, F., {et~al.} 2022, \bibinfo{title}{{Deep Unsupervised Domain Adaptation: A Review of Recent Advances and Perspectives},} arXiv e-prints, arXiv:2208.07422, \dodoi{10.48550/arXiv.2208.07422}

\bibitem[{D. {L{\'o}pez-Cano} {et~al.}(2024){L{\'o}pez-Cano}, {St{\"u}cker}, {Pellejero Iba{\~n}ez}, {Angulo}, \& {Franco-Barranco}}]{2024A&A...685A..37L}
{L{\'o}pez-Cano}, D., {St{\"u}cker}, J., {Pellejero Iba{\~n}ez}, M., {Angulo}, R.~E., \& {Franco-Barranco}, D. 2024, \bibinfo{title}{{Characterizing structure formation through instance segmentation},} \aap, 685, A37, \dodoi{10.1051/0004-6361/202348965}

\bibitem[{G. {Mart{\'\i}nez-Solaeche} {et~al.}(2023){Mart{\'\i}nez-Solaeche}, {Queiroz}, {Gonz{\'a}lez Delgado}, {Rodrigues}, {Garc{\'\i}a-Benito}, {P{\'e}rez-R{\`a}fols}, {Raul Abramo}, {D{\'\i}az-Garc{\'\i}a}, {Pieri}, {Chaves-Montero}, {Hern{\'a}n-Caballero}, {Rodr{\'\i}guez-Mart{\'\i}n}, {Bonoli}, {Morrison}, {M{\'a}rquez}, {V{\'\i}lchez}, {Fern{\'a}ndez-Ontiveros}, {Marra}, {Alcaniz}, {Benitez}, {Cenarro}, {Crist{\'o}bal-Hornillos}, {Dupke}, {Ederoclite}, {L{\'o}pez-Sanjuan}, {Mar{\'\i}n-Franch}, {Mendes de Oliveira}, {Moles}, {Sodr{\'e}}, {Taylor}, {Varela}, \& {V{\'a}zquez Rami{\'o}}}]{2023A&A...673A.103M}
{Mart{\'\i}nez-Solaeche}, G., {Queiroz}, C., {Gonz{\'a}lez Delgado}, R.~M., {et~al.} 2023, \bibinfo{title}{{The miniJPAS survey quasar selection. III. Classification with artificial neural networks and hybridisation},} \aap, 673, A103, \dodoi{10.1051/0004-6361/202245750}

\bibitem[{S. {Pandya} {et~al.}(2025){Pandya}, {Patel}, {Nord}, {Walmsley}, \& {{\'C}iprijanovi{\'c}}}]{2025MLS&T...6c5032P}
{Pandya}, S., {Patel}, P., {Nord}, B.~D., {Walmsley}, M., \& {{\'C}iprijanovi{\'c}}, A. 2025, \bibinfo{title}{{SIDDA: SInkhorn Dynamic Domain Adaptation for image classification with equivariant neural networks},} Machine Learning: Science and Technology, 6, 035032, \dodoi{10.1088/2632-2153/adf701}

\bibitem[{H. {Parul} {et~al.}(2025){Parul}, {Gleyzer}, {Reddy}, \& {Toomey}}]{2025ApJ...990...47P}
{Parul}, H., {Gleyzer}, S., {Reddy}, P., \& {Toomey}, M.~W. 2025, \bibinfo{title}{{Domain Adaptation in Application to Gravitational Lens Finding},} \apj, 990, 47, \dodoi{10.3847/1538-4357/adee16}

\bibitem[{A. {Paszke} {et~al.}(2019){Paszke}, {Gross}, {Massa}, {Lerer}, {Bradbury}, {Chanan}, {Killeen}, {Lin}, {Gimelshein}, {Antiga}, {Desmaison}, {K{\"o}pf}, {Yang}, {DeVito}, {Raison}, {Tejani}, {Chilamkurthy}, {Steiner}, {Fang}, {Bai}, \& {Chintala}}]{2019arXiv191201703P}
{Paszke}, A., {Gross}, S., {Massa}, F., {et~al.} 2019, \bibinfo{title}{{PyTorch: An Imperative Style, High-Performance Deep Learning Library},} arXiv e-prints, arXiv:1912.01703, \dodoi{10.48550/arXiv.1912.01703}

\bibitem[{I. {P{\'e}rez-R{\`a}fols} {et~al.}(2023){P{\'e}rez-R{\`a}fols}, {Abramo}, {Mart{\'\i}nez-Solaeche}, {Pieri}, {Queiroz}, {Rodrigues}, {Bonoli}, {Chaves-Montero}, {Morrison}, {Alcaniz}, {Benitez}, {Carneiro}, {Cenarro}, {Crist{\'o}bal-Hornillos}, {Dupke}, {Ederoclite}, {Gonz{\'a}lez Delgado}, {Hern{\'a}n-Caballero}, {L{\'o}pez-Sanjuan}, {Mar{\'\i}n-Franch}, {Marra}, {Mendes de Oliveira}, {Moles}, {Sodr{\'e}}, {Taylor}, {Varela}, \& {V{\'a}zquez Rami{\'o}}}]{2023A&A...678A.144P}
{P{\'e}rez-R{\`a}fols}, I., {Abramo}, L.~R., {Mart{\'\i}nez-Solaeche}, G., {et~al.} 2023, \bibinfo{title}{{The miniJPAS survey quasar selection. IV. Classification and redshift estimation with SQUEzE},} \aap, 678, A144, \dodoi{10.1051/0004-6361/202347488}

\bibitem[{I. {P{\'e}rez-R{\`a}fols} {et~al.}(2025){P{\'e}rez-R{\`a}fols}, {Abramo}, {Mart{\'\i}nez-Solaeche}, {Rodrigues}, {Pieri}, {Burjal{\`e}s-del-Amo}, {Escol{\`a}-Gallinat}, {Ferr{\'e}-Abad}, {Isern-Vizoso}, {Alcaniz}, {Benitez}, {Bonoli}, {Carneiro}, {Cenarro}, {Crist{\'o}bal-Hornillos}, {Dupke}, {Ederoclite}, {Mar{\'\i}a Gonz{\'a}lez Delgado}, {Gurung-Lopez}, {Hern{\'a}n-Caballero}, {Hern{\'a}ndez-Monteagudo}, {L{\'o}pez-Sanjuan}, {Mar{\'\i}n-Franch}, {Marra}, {Mendes de Oliveira}, {Moles}, {Sodr{\'e}}, {Taylor}, {Varela}, \& {V{\'a}zquez Rami{\'o}}}]{2025arXiv250711380P}
{P{\'e}rez-R{\`a}fols}, I., {Abramo}, L.~R., {Mart{\'\i}nez-Solaeche}, G., {et~al.} 2025, \bibinfo{title}{{The miniJPAS survey quasar selection V: combined algorithm},} arXiv e-prints, arXiv:2507.11380, \dodoi{10.48550/arXiv.2507.11380}

\bibitem[{M.~M. {Pieri} {et~al.}(2016){Pieri}, {Bonoli}, {Chaves-Montero}, {P{\^a}ris}, {Fumagalli}, {Bolton}, {Viel}, {Noterdaeme}, {Miralda-Escud{\'e}}, {Busca}, {Rahmani}, {Peroux}, {Font-Ribera}, \& {Trager}}]{2016sf2a.conf..259P}
{Pieri}, M.~M., {Bonoli}, S., {Chaves-Montero}, J., {et~al.} 2016, \bibinfo{title}{{WEAVE-QSO: A Massive Intergalactic Medium Survey for the William Herschel Telescope},} in SF2A-2016: Proceedings of the Annual meeting of the French Society of Astronomy and Astrophysics, ed. C.~{Reyl{\'e}}, J.~{Richard}, L.~{Cambr{\'e}sy}, M.~{Deleuil}, E.~{P{\'e}contal}, L.~{Tresse}, \& I.~{Vauglin}, 259--266, \dodoi{10.48550/arXiv.1611.09388}

\bibitem[{S. {Pierre} {et~al.}(2025){Pierre}, {R{\'e}galdo-Saint Blancard}, {Hahn}, \& {Eickenberg}}]{2025arXiv250703086P}
{Pierre}, S., {R{\'e}galdo-Saint Blancard}, B., {Hahn}, C., \& {Eickenberg}, M. 2025, \bibinfo{title}{{Mitigating Model Misspecification in Simulation-Based Inference for Galaxy Clustering},} arXiv e-prints, arXiv:2507.03086, \dodoi{10.48550/arXiv.2507.03086}

\bibitem[{D. {Potter} {et~al.}(2017){Potter}, {Stadel}, \& {Teyssier}}]{2017ComAC...4....2P}
{Potter}, D., {Stadel}, J., \& {Teyssier}, R. 2017, \bibinfo{title}{{PKDGRAV3: beyond trillion particle cosmological simulations for the next era of galaxy surveys},} Computational Astrophysics and Cosmology, 4, 2, \dodoi{10.1186/s40668-017-0021-1}

\bibitem[{C. {Queiroz} {et~al.}(2023){Queiroz}, {Abramo}, {Rodrigues}, {P{\'e}rez-R{\`a}fols}, {Mart{\'\i}nez-Solaeche}, {Hern{\'a}n-Caballero}, {Hern{\'a}ndez-Monteagudo}, {Lumbreras-Calle}, {Pieri}, {Morrison}, {Bonoli}, {Chaves-Montero}, {Chies-Santos}, {D{\'\i}az-Garc{\'\i}a}, {Fernandez-Soto}, {Gonz{\'a}lez Delgado}, {Alcaniz}, {Ben{\'\i}tez}, {Cenarro}, {Civera}, {Dupke}, {Ederoclite}, {L{\'o}pez-Sanjuan}, {Mar{\'\i}n-Franch}, {Mendes de Oliveira}, {Moles}, {Muniesa}, {Sodr{\'e}}, {Taylor}, {Varela}, \& {V{\'a}zquez Rami{\'o}}}]{2023MNRAS.520.3476Q}
{Queiroz}, C., {Abramo}, L.~R., {Rodrigues}, N. V.~N., {et~al.} 2023, \bibinfo{title}{{The miniJPAS survey quasar selection - I. Mock catalogues for classification},} \mnras, 520, 3476, \dodoi{10.1093/mnras/stac2962}

\bibitem[{C. {Rampf} {et~al.}(2025){Rampf}, {List}, \& {Hahn}}]{2025JCAP...02..020R}
{Rampf}, C., {List}, F., \& {Hahn}, O. 2025, \bibinfo{title}{{BULLFROG: multi-step perturbation theory as a time integrator for cosmological simulations},} \jcap, 2025, 020, \dodoi{10.1088/1475-7516/2025/02/020}

\bibitem[{N.~V.~N. {Rodrigues} {et~al.}(2023){Rodrigues}, {Raul Abramo}, {Queiroz}, {Mart{\'\i}nez-Solaeche}, {P{\'e}rez-R{\`a}fols}, {Bonoli}, {Chaves-Montero}, {Pieri}, {Gonz{\'a}lez Delgado}, {Morrison}, {Marra}, {M{\'a}rquez}, {Hern{\'a}n-Caballero}, {D{\'\i}az-Garc{\'\i}a}, {Ben{\'\i}tez}, {Cenarro}, {Dupke}, {Ederoclite}, {L{\'o}pez-Sanjuan}, {Mar{\'\i}n-Franch}, {Mendes de Oliveira}, {Moles}, {Sodr{\'e}}, {Varela}, {V{\'a}zquez Rami{\'o}}, \& {Taylor}}]{2023MNRAS.520.3494R}
{Rodrigues}, N. V.~N., {Raul Abramo}, L., {Queiroz}, C., {et~al.} 2023, \bibinfo{title}{{The miniJPAS survey quasar selection - II. Machine learning classification with photometric measurements and uncertainties},} \mnras, 520, 3494, \dodoi{10.1093/mnras/stac2836}

\bibitem[{A. {Roncoli} {et~al.}(2023){Roncoli}, {{\'C}iprijanovi{\'c}}, {Voetberg}, {Villaescusa-Navarro}, \& {Nord}}]{2023arXiv231101588R}
{Roncoli}, A., {{\'C}iprijanovi{\'c}}, A., {Voetberg}, M., {Villaescusa-Navarro}, F., \& {Nord}, B. 2023, \bibinfo{title}{{Domain Adaptive Graph Neural Networks for Constraining Cosmological Parameters Across Multiple Data Sets},} arXiv e-prints, arXiv:2311.01588, \dodoi{10.48550/arXiv.2311.01588}

\bibitem[{J. {Schaye} {et~al.}(2023){Schaye}, {Kugel}, {Schaller}, {Helly}, {Braspenning}, {Elbers}, {McCarthy}, {van Daalen}, {Vandenbroucke}, {Frenk}, {Kwan}, {Salcido}, {Bah{\'e}}, {Borrow}, {Chaikin}, {Hahn}, {Hu{\v{s}}ko}, {Jenkins}, {Lacey}, \& {Nobels}}]{2023MNRAS.526.4978S}
{Schaye}, J., {Kugel}, R., {Schaller}, M., {et~al.} 2023, \bibinfo{title}{{The FLAMINGO project: cosmological hydrodynamical simulations for large-scale structure and galaxy cluster surveys},} \mnras, 526, 4978, \dodoi{10.1093/mnras/stad2419}

\bibitem[{V. {Springel} {et~al.}(2021){Springel}, {Pakmor}, {Zier}, \& {Reinecke}}]{2021MNRAS.506.2871S}
{Springel}, V., {Pakmor}, R., {Zier}, O., \& {Reinecke}, M. 2021, \bibinfo{title}{{Simulating cosmic structure formation with the GADGET-4 code},} \mnras, 506, 2871, \dodoi{10.1093/mnras/stab1855}

\bibitem[{P. {Swierc} {et~al.}(2024){Swierc}, {Tamargo-Arizmendi}, {{\'C}iprijanovi{\'c}}, \& {Nord}}]{2024arXiv241016347S}
{Swierc}, P., {Tamargo-Arizmendi}, M., {{\'C}iprijanovi{\'c}}, A., \& {Nord}, B.~D. 2024, \bibinfo{title}{{Domain-Adaptive Neural Posterior Estimation for Strong Gravitational Lens Analysis},} arXiv e-prints, arXiv:2410.16347, \dodoi{10.48550/arXiv.2410.16347}

\bibitem[{P. {Swierc} {et~al.}(2023){Swierc}, {Zhao}, {{\'C}iprijanovi{\'c}}, \& {Nord}}]{2023arXiv231117238S}
{Swierc}, P., {Zhao}, M., {{\'C}iprijanovi{\'c}}, A., \& {Nord}, B. 2023, \bibinfo{title}{{Domain Adaptation for Measurements of Strong Gravitational Lenses},} arXiv e-prints, arXiv:2311.17238, \dodoi{10.48550/arXiv.2311.17238}

\bibitem[{E. {Tzeng} {et~al.}(2017){Tzeng}, {Hoffman}, {Saenko}, \& {Darrell}}]{2017arXiv170205464T}
{Tzeng}, E., {Hoffman}, J., {Saenko}, K., \& {Darrell}, T. 2017, \bibinfo{title}{{Adversarial Discriminative Domain Adaptation},} arXiv e-prints, arXiv:1702.05464, \dodoi{10.48550/arXiv.1702.05464}

\bibitem[{S. {van der Walt} {et~al.}(2011){van der Walt}, {Colbert}, \& {Varoquaux}}]{Numpy-vanDerWalt11}
{van der Walt}, S., {Colbert}, S.~C., \& {Varoquaux}, G. 2011, \bibinfo{title}{The NumPy Array: A Structure for Efficient Numerical Computation,} Computing in Science Engineering, 13, 22, \dodoi{10.1109/MCSE.2011.37}

\bibitem[{J. {Vega-Ferrero} {et~al.}(2024){Vega-Ferrero}, {Huertas-Company}, {Costantin}, {P{\'e}rez-Gonz{\'a}lez}, {Sarmiento}, {Kartaltepe}, {Pillepich}, {Bagley}, {Finkelstein}, {McGrath}, {Knapen}, {Arrabal Haro}, {Bell}, {Buitrago}, {Calabr{\`o}}, {Dekel}, {Dickinson}, {Dom{\'\i}nguez S{\'a}nchez}, {Elbaz}, {Ferguson}, {Giavalisco}, {Holwerda}, {Kocesvski}, {Koekemoer}, {Pandya}, {Papovich}, {Pirzkal}, {Primack}, \& {Yung}}]{2024ApJ...961...51V}
{Vega-Ferrero}, J., {Huertas-Company}, M., {Costantin}, L., {et~al.} 2024, \bibinfo{title}{{On the Nature of Disks at High Redshift Seen by JWST/CEERS with Contrastive Learning and Cosmological Simulations},} \apj, 961, 51, \dodoi{10.3847/1538-4357/ad05bb}

\bibitem[{M. {Vogelsberger} {et~al.}(2014){Vogelsberger}, {Genel}, {Springel}, {Torrey}, {Sijacki}, {Xu}, {Snyder}, {Nelson}, \& {Hernquist}}]{2014MNRAS.444.1518V}
{Vogelsberger}, M., {Genel}, S., {Springel}, V., {et~al.} 2014, \bibinfo{title}{{Introducing the Illustris Project: simulating the coevolution of dark and visible matter in the Universe},} \mnras, 444, 1518, \dodoi{10.1093/mnras/stu1536}

\bibitem[{R. {von Marttens} {et~al.}(2024){von Marttens}, {Marra}, {Quartin}, {Casarini}, {Baqui}, {Alvarez-Candal}, {Galindo-Guil}, {Fern{\'a}ndez-Ontiveros}, {del Pino}, {D{\'\i}az-Garc{\'\i}a}, {L{\'o}pez-Sanjuan}, {Alcaniz}, {Angulo}, {Cenarro}, {Crist{\'o}bal-Hornillos}, {Dupke}, {Ederoclite}, {Hern{\'a}ndez-Monteagudo}, {Mar{\'\i}n-Franch}, {Moles}, {Sodr{\'e}}, {Varela}, \& {V{\'a}zquez Rami{\'o}}}]{2024MNRAS.527.3347V}
{von Marttens}, R., {Marra}, V., {Quartin}, M., {et~al.} 2024, \bibinfo{title}{{J-PLUS: galaxy-star-quasar classification for DR3},} \mnras, 527, 3347, \dodoi{10.1093/mnras/stad3373}

\bibitem[{J. {Wang} {et~al.}(2021){Wang}, {Lan}, {Liu}, {Ouyang}, {Qin}, {Lu}, {Chen}, {Zeng}, \& {Yu}}]{2021arXiv210303097W}
{Wang}, J., {Lan}, C., {Liu}, C., {et~al.} 2021, \bibinfo{title}{{Generalizing to Unseen Domains: A Survey on Domain Generalization},} arXiv e-prints, arXiv:2103.03097, \dodoi{10.48550/arXiv.2103.03097}

\bibitem[{Y.-C. {Yu} \& H.-T. {Lin}(2023){Yu} \& {Lin}}]{2023arXiv230202335Y}
{Yu}, Y.-C., \& {Lin}, H.-T. 2023, \bibinfo{title}{{Semi-Supervised Domain Adaptation with Source Label Adaptation},} arXiv e-prints, arXiv:2302.02335, \dodoi{10.48550/arXiv.2302.02335}

\bibitem[{K. {Zhou} {et~al.}(2021){Zhou}, {Liu}, {Qiao}, {Xiang}, \& {Change Loy}}]{2021arXiv210302503Z}
{Zhou}, K., {Liu}, Z., {Qiao}, Y., {Xiang}, T., \& {Change Loy}, C. 2021, \bibinfo{title}{{Domain Generalization: A Survey},} arXiv e-prints, arXiv:2103.02503, \dodoi{10.48550/arXiv.2103.02503}

\end{thebibliography}
\bibliographystyle{aasjournalv7}



\end{document}